\documentclass[prb,preprint]{revtex4-1}

\usepackage[x11names]{xcolor}

\usepackage{amsmath} 
\usepackage{amsfonts} 
\usepackage{graphicx} 
\usepackage{amssymb}
\usepackage{subcaption}
\usepackage{tikz}
\usepackage{tikzsymbols}
\usetikzlibrary{snakes}
\usetikzlibrary{decorations.pathmorphing,math,calc,3d,shapes}
\usetikzlibrary{arrows.meta,arrows,angles,quotes,patterns}
\usepackage{tikz-3dplot}
\usepackage{pgfplots}
\usepackage{mathtools}
\def\centerarc[#1](#2)(#3:#4:#5){
    \draw[#1]([shift=(#3:#5)]#2) arc (#3:#4:#5);
    }
\tikzset{
    partial ellipse/.style args={#1:#2:#3}{
        insert path={+ (#1:#3) arc (#1:#2:#3)}
    }
}
\usetikzlibrary{decorations.markings}
\tikzset{
    set arrow inside/.code={\pgfqkeys{/tikz/arrow inside}{#1}},
    set arrow inside={end/.initial=>, opt/.initial=},
    /pgf/decoration/Mark/.style={
        mark/.expanded=at position #1 with
        {
            \noexpand\arrow[\pgfkeysvalueof{/tikz/arrow inside/opt}]{\pgfkeysvalueof{/tikz/arrow inside/end}}
        }
    },
    arrow inside/.style 2 args={
        set arrow inside={#1},
        postaction={
            decorate,decoration={
                markings,Mark/.list={#2}
            }
        }
    },
}
\tikzstyle{arrowmid}[0.5]=[decoration=
{markings,
mark=at position #1 with {\arrow{>}}},
postaction={decorate}]

\newcommand\be{\begin{equation}}
\newcommand\ee{\end{equation}}

\newcommand\bse{\begin{subequations}}
\newcommand\ese{\end{subequations}}

\begin{document}
\title{Work Done by Sliding Friction}
\author{J. David Brown}
\email{david\_brown@ncsu.edu}
\affiliation{Department of Physics and Astronomy\\
North Carolina State University, Raleigh, NC 27695}
\date{\today}

\begin{abstract}  
Friction does work when two objects in contact  slide past one another. 
This process is studied using a simple numerical model consisting of two flexible asperites, one for each object. The force of friction 
is modeled as a conservative electric force that can be either attractive or repulsive. Because the asperites are flexible, not rigid, 
the conservative friction force generates thermal energy and  can 
decrease or increase the objects' translational kinetic energies.  
The concept of pseudowork plays an important 
role throughout the analysis. The difference between work and pseudowork is equal to the change in internal energy. 
The numerical model is used to illuminate the reasons why work is typically greater than pseudowork. 
For a real physical system, this is the source of thermal energy.  
The processes discussed here fall into 
two categories: forced sliding  and free sliding. With forced sliding, applied forces keep the objects moving with constant velocities. With free sliding, 
the only force on one of the objects (in the direction of motion) is friction.  
\end{abstract}
\maketitle

\section{Introduction}
Textbook treatments of sliding friction  typically focus on the empirical laws developed in separate works 
by da Vinci \cite{Hutchings,MacCurdy}, Amontons \cite{Amontons}, and Coulomb,\cite{Coulomb} which state  that the force of 
friction is proportional to  the normal force, is  independent 
of the area of contact between surfaces, and is independent of the relative velocity between surfaces. 

These simple laws are suitable for an elementary analysis of the forces  
acting on objects as they slide past one another.  They are not adequate for explaining the transfer of energy. 
This issue is highlighted in the textbook ``Matter and Interactions,"\cite{M+I} where the authors discuss the following simple thought 
experiment.\cite{SherwoodBernard}  A block slides across a table, from left to right,  in response to applied forces $\vec f_a$  and a 
friction force $\vec f_f$. (The applied forces might include gravity and a normal force, but these typically cancel.) Since the 
velocity of the  block is constant, the net force must be zero, $\vec f_a + \vec f_f = 0$. Naively, the work done by applied forces is $w_a = \vec f_a \cdot \Delta \vec r$
and the work done by friction is $w_f = \vec f_f \cdot \Delta \vec r$, where 
$\Delta \vec r$ is the displacement of the block.  
Since $\vec f_f = -\vec f_a$, this implies that the net work done on the block 
vanishes, $w_a + w_f = 0$.  Of course this conclusion must be wrong because 
the temperature of the block rises. If no work is done on the block, where does the increase in thermal energy come from? 

As discussed in ``Matter and Interactions,"  
contact between the block and table occurs at asperites, high points on each surface.  
The asperites are not rigid, rather,  they deform as they slip past one another. This leads to an effective displacement for the friction force that differs from
 the displacement of the block's center of mass.  The positive work done by the applied force is larger in magnitude than the negative work done by friction.   
 The net work is positive, and this is the source of thermal energy.  
 
 This situation, in which applied forces keep the block moving with constant velocity, will be referred to as ``forced sliding."  We also consider 
 ``free sliding," in which the block slides across the table without any applied force in the direction of motion. Only friction acts 
 in the  direction of motion. 

In this paper we analyze the thought experiment from ``Matter and Interactions," and at the same time address another puzzle:  
At a microscopic level,  the interaction between block and table, or between asperites,  arises due to electric forces between electrons and protons.
 But electric forces are conservative, that is, 
derivable from a potential.  How does friction, which is often characterized as a nonconservative force, arise from a fundamentally conservative force? 

To help illustrate this puzzle, consider the following thought experiment. A block freely slides 
from one end of a table to the other. Asperites in the table exert conservative forces on asperites in the block. 
Now imagine extending the path of the block to $\pm\infty$, and then closing 
the path  by adding a semicircle at infinity. 
If the block is carried around the loop, the work done on an asperite in the block  by an asperite in the table 
is zero. After all, this is a defining property of a 
conservative force.  Moreover, no work is done along the added path 
elements because  the interaction force vanishes when the asperites are far apart. 
It seems  that the work done by friction must vanish as the upper block freely slides along the table. 

Of course this reasoning is wrong.  With free sliding on a stationary table, the block slows down and it's kinetic energy decreases. 
The error in reasoning comes from treating the asperites as rigid objects. 
Although the friction force (the force that the table exerts on the block) is fundamentally conservative, this force can do nonzero 
work because the asperites deform. The force fields created by the charges in the asperites are not static, and this allows the conservative 
force of friction to do work.

Historically, friction was viewed as a simple consequence of surface roughness. 
This explanation was ruled out by the 1950's and replaced with the concept of molecular adhesion.\cite{BowdenTabor} 
The essential role of deformation in the description of friction was first recognized by Tomlinson,\cite{Tomlinson} and has been 
rediscovered many times thereafter.\cite{Sokoloff,GlosliMcClelland}  We now understand that 
friction occurs when interacting asperites are set into motion, and this motion gives rise to 
lattice vibrations (phonons) that eventually dissipate into thermal energy.\cite{Krim, Ringlein, Vanossi, Corpuz} 

In this article we present a simple  model for  sliding friction with fundamentally conservative forces. 
We do not address the laws of da Vinci, Amontons and Coulomb. This model  is simply intended to provide insight into the mechanism 
by which energy is transferred in the presence of sliding friction.  
The model, described in Sec.~\ref{Sec:model}, consists of two asperites that slide past one another. 
Each asperite is modeled as a pendulum connected to a bar with a torsion spring.  
The pendulum bobs at the ends of each asperite interact via  simple electric
attraction or repulsion. The numerical model can be extended easily by replacing the interaction with a more realistic 
intermolecular potential, such as the  Morse\cite{Morse} or Lennard--Jones potentials.\cite{Jones}
The equations of motion for the model asperites are derived in Sec.~\ref{Sec:EOM}. 

In Sec.~\ref{Sec:work} we discuss work and energy. The work done by a force is  the vector product of the force and the
displacement of the point of application of the force. The work--energy principle (first law of thermodynamics) tells us that the 
net work done on a system is equal to the change in energy of the system. 
We also define the  pseudowork\cite{Erlichson,Penchina,Sherwood} done by a force as the vector product of the force and the 
displacement of the system's center of mass. 
The pseudowork--energy principle tells us that the  net pseudowork done on a system is equal to the change in  translational kinetic energy
of the system. Finally, the invariant first law of thermodynamics\cite{SherwoodBernard} (IFLT) says that the change in internal energy (including thermal energy) 
is equal to the difference between work and pseudowork. Since friction increases thermal energy, 
 work must be greater than  pseudowork. 

In Sec.~\ref{Sec:forces} we discuss our choices for the applied forces, as well as the interaction forces between asperites, for both forced and free sliding. 
The numerical implementation is discussed briefly in Sec.~\ref{Sec:numerics}. The numerical code is included as supplementary material, and is 
also freely available at glowscript.org.\cite{glowscript}
 
The numerical results  are presented in Sec.~\ref{Sec:ExamplesResults} where we address (A) 
forced sliding with repulsive interaction, (B) forced sliding with attractive interaction, (C) free sliding on a fixed surface with 
repulsive interaction, (D) free sliding 
on a fixed surface with attractive interaction, and (E) free sliding from rest on a moving surface with both repulsive and attractive interactions. 
The key question is the following: Why should we expect the  work to be greater than the  pseudowork, for both  
attractive and repulsive interactions? Moreover, 
for free sliding on a stationary surface, the net pseudowork must be negative for the block to slow down. 
For free sliding from rest on a moving surface, the net pseudowork must be positive for the block to speed up.
Why should we expect the pseudowork to be negative  when the surface is stationary, 
and positive  when the block starts from rest and the surface is moving? 

Results are discussed in Sec.~\ref{Sec:discuss}, where we 
propose a conceptual--level explanation for 
the increase in thermal energy in the presence of sliding friction.  
We also point out the limitations of this work: The processes considered here are time--reversible, so 
the  internal (thermal) energy can either increase or decrease. Ultimately, the direction of energy transfer depends on 
initial conditions. A more complete understanding of  work done by friction  would require a statistical analysis.

\section{The model}\label{Sec:model}
\begin{figure}[h!]
\centering
\begin{tikzpicture}[scale=1.5, decoration={coil,amplitude=0.5mm,segment length=0.5mm}]
	\draw[thick, rotate=15, fill=blue!20] (-2.5,-0.15)--(2.5,-0.15)--(2.5,0.15)--(-2.5,0.15)--cycle;  
	\draw [thick, fill=blue!20] (1.65,-2.85) circle (0.3);                                                          
	\draw[thick, rotate=30, fill=yellow!20] (-0.05,0)--(-0.05,-3)--(0.05,-3)--(0.05,0)--cycle;   
	\draw[dashed, ->] (-2,-3.4) -- (-0.5,-3.4) node[right] {$x$};     
	\draw[dashed, ->] (-2,-3.4) -- (-2,-2.2) node[left] {$y$};    
	\draw[fill] (0,0) circle (0.05);                                         
	\draw[thick, fill=red!40] (-60:1.4) circle (0.07)          ;   
	\centerarc[decorate](0,0)(189:299:0.3);    
	\centerarc[decorate](0,0)(301:375:0.3);    
	%
	\draw[thick, ->] (1.65,-2.85) --+ (1.6,1.3) node[right] {$\vec f_1$};
	\draw[thick, ->] (0,0) --+(-1.3,1.1) node[left] {$\vec f_2$};
	\draw[thick, ->] (2.42,0.63) --+(0.4,1.0) node[right] {$\vec f_R$};
	\draw[thick, ->] (-2.42,-0.63) --+(-0.2,-0.9) node[left] {$\vec f_L$};
	%
	\draw[thick, ->] (-2,-3.4) -- (0.62,-1.25);       
	\node at (-0.3,-1.7)  {$\vec r_{\rm cm}$};
	\draw[thick, ->] (-2,-3.4) -- (1.65,-2.85) node[midway, above] {$\vec r_1$};       
	\draw[thick, ->] (-2,-3.4) -- (-0.05,-0.05) node[midway, left] {$\vec r_2$};           
	\draw[thick, ->] (0,0) -- (2.42,0.63);              
	\node at (2.7,0.65) {$\vec r_R$};
	\draw[thick, ->] (0,0) -- (-2.42,-0.63);            
	\node at (-2.7,-0.65) {$\vec r_L$};
	%
	\draw[dashed] (0,0) -- (2.7,0);     
	\draw[dashed] (0,-1.3) -- (0,0);    
	\centerarc[](0,0)(270:300:0.95);    
	\node at (285:1.1) {$\theta$};
	\centerarc[](0,0)(0:15:2.0);          
	\node at (5.2:2.15) {$\phi$};
	\node at (0.65,-0.6) {$\ell_2$}; 
	\node at (1.05,-1.2) {cm}; 
	\node at (1.4,-1.9) {$\ell_1$}; 
	\node at (2.24,-2.85) {$m_1$};
	\node at (0.15,0.5) {$m_2,\ i_2$}; 
	\node at (-1.4,0) {$a/2$};
	\node at (1.2,0.7) {$a/2$}; 
\end{tikzpicture}
\caption{Each asperite consists of a bar and a pendulum, connected with a torsion spring.}
\label{UpperBlockFig}
\end{figure}
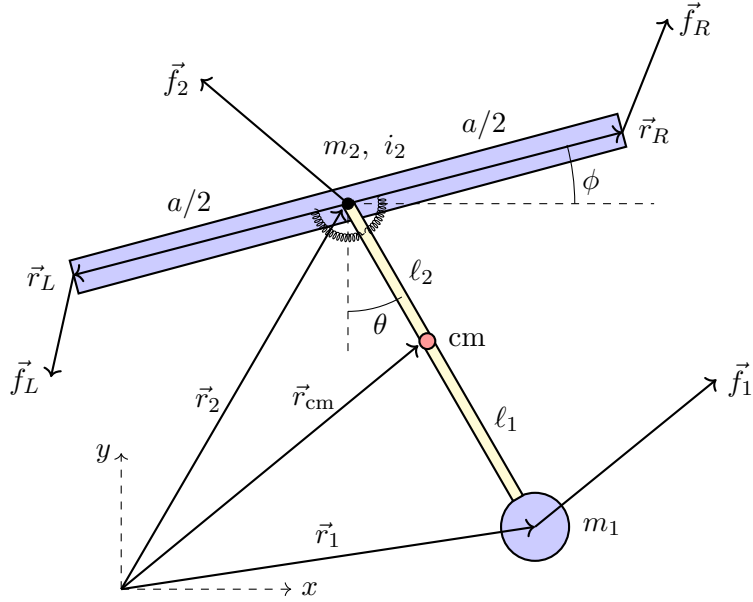

The model replaces the block with a single asperite, the ``upper asperite," and replaces the table with a single 
``lower asperite."  The upper asperite is shown in Fig.~\ref{UpperBlockFig}. 
It consists of a uniform bar and a pendulum that swings from the center of the bar.  The   asperite's center of mass 
lies on the pendulum rod, and is shown as a red dot. 
The bar has length $a$, mass $m_2$ and moment of inertia (about its center of mass) 
$i_2$. The mass of the pendulum bob is $m_1$. 
The pendulum rod is massless, and has length $\ell = \ell_1 + \ell_2$.  

The pendulum is configured with a torsion spring, represented by the curved springs in the figure. 
For the upper asperite, the equilibrium position of the torsion spring  is  $\theta = \phi$. 

The  lower asperite is identical to the  upper asperite, apart from the equilibrium position of the torsion spring.  
For the lower asperite we use capital letters for lengths,  angles and parameters. For example, the 
pendulum rod has length $L = L_1 + L_2$ and it makes an angle of $\Theta$ with respect to the vertical direction. 
The equilibrium position for the torsion spring on the lower asperite is $\Theta = \Phi + \pi$. 

Position vectors and  forces on the upper asperite are show in Fig.~\ref{UpperBlockFig}. 
The position of the center of mass  is  $\vec r_{\rm cm}$. Position vectors for the pendulum bob and the center of the bar are
(angle brackets are used to surround the Cartesian coordinate components of a vector)
\bse
\begin{align}
	\vec r_1 &= \vec r_{\rm cm} + \ell_1 \langle  \sin\theta, - \cos\theta, 0 \rangle \ ,\\
	\vec r_2 &= \vec r_{\rm cm} + \ell_2 \langle - \sin\theta,  \cos\theta, 0 \rangle \ ,
\end{align}
\ese
where $\ell_1 = \ell m_2 /(m_1 + m_2)$ and $\ell_2 = \ell m_1/(m_1 + m_2)$.  The vectors  
\bse
\begin{align}
	\vec r_L &= (a/2) \langle -\cos\phi, -\sin\phi, 0 \rangle   \ ,\\
	\vec r_R &=(a/2) \langle \cos\phi, \sin\phi,0 \rangle 
\end{align}
\ese
define the positions of the left and right ends of the bar relative to the bar's center of mass.

The force that the lower asperite exerts on the upper asperite is $\vec f_1$.  That is, $\vec f_1$ is the 
force of friction. The force that the upper asperite exerts on the lower asperite is $\vec F_1$. 
By Newton's third law, $\vec F_1 = -\vec f_1$. 

The remaining forces $\vec f_2$, $\vec f_L$ and $\vec f_R$ are applied forces. These are meant to represent the forces 
that the bulk material of the block exerts on the upper asperite to keep it in place. Note that the gravitational force on the bar can be absorbed into these 
applied forces. The gravitational force on the pendulum bob is ignored---we imagine it to be significantly smaller than the friction force $\vec f_1$
and the torsion spring force. 

This model only addresses the interaction of two asperites. The internal energy of each asperite is represented by the motion of the 
pendulum and the stretch or compression of the torsion springs. For a more complete model of a block and table, one might add (many) 
harmonic oscillators coupled to each bar. After the asperites interact, their internal energies would dissipate into the coupled oscillators.  
This would represent the transfer of energy from the flexible asperites into thermal energy of the bulk material. 

\section{Equations of motion}\label{Sec:EOM}
Consider the upper asperite. The torsion spring exerts a torque $\langle 0, 0, -\kappa(\theta - \phi) \rangle$ 
 on the pendulum rod, and a torque $\langle 0, 0, \kappa(\theta - \phi) \rangle$ 
 on the bar. For the moment, consider the bar as the system. The torque--angular momentum principle  tells us that the 
 sum of external torques about the bar's center of mass equals the rate 
 of change of angular  momentum of the bar: 
 \be\label{AMPforbar}
 	\langle 0, 0, \kappa(\theta - \phi) \rangle + \vec r_R \times \vec f_R + \vec r_L \times \vec f_L = \langle 0, 0, i_2 \ddot\phi \rangle \ .
\ee 
Dots denote time derivatives. 

Now consider the entire upper asperite (bar plus pendulum) as the system. By Newton's second law, 
the sum of external forces equals the product of mass and acceleration of the center of mass: 
\be\label{NSL}
	\vec f_1 + \vec f_2 + \vec f_L + \vec f_R = (m_1 + m_2) \ddot{\vec r}_{\rm cm} 
\ee
The torque--angular momentum principle is
$ \sum \vec\tau_{\rm cm}  = d\vec L_{\rm cm}/dt $, where 
\be
 	\sum \vec\tau_{\rm cm}  = (\vec r_1 - \vec r_{\rm  cm})\times \vec f_1 + (\vec r_2 - \vec r_{\rm cm}) \times \vec f_2 
	+ (\vec r_2 + \vec r_R - \vec r_{\rm cm}) \times \vec f_R +  (\vec r_2 + \vec r_L - \vec r_{\rm cm}) \times \vec f_L   
\ee
is the net torque on the system (the upper asperite) about the system's center of mass, and 
\be
	\vec L_{\rm cm} = m_1 (\vec r_1 - \vec r_{\rm cm})\times (\dot{\vec r}_1 - \dot{\vec r}_{\rm cm}) + 
	m_2 (\vec r_2 - \vec r_{\rm cm})\times (\dot{\vec r}_2 - \dot{\vec r}_{\rm cm}) + \langle 0, 0, i_2 \dot\phi \rangle   
\ee
is the angular momentum of the system about its center of mass. 
The first two terms are the translational angular momenta of the pendulum bob and bar, respectively. The third term is the 
rotational (spin) angular momentum of the bar about its own center of mass. 

Solving the above equations  for the accelerations, we find 
\bse\label{accelerations}
\begin{align}
	\ddot{\vec r}_{\rm cm} &= \frac{1}{m_1 + m_2} \left[ \vec f_1 + \vec f_2 + \vec f_L + \vec f_R \right] \ ,\\
	\ddot\theta &= \frac{1}{m_1 l_1^2 + m_2 l_2^2} \left[  ((\vec r_1 - \vec r_{\rm cm})\times \vec f_1)_z 
	+ ((\vec r_2 - \vec r_{\rm cm})\times (\vec f_2 + \vec f_L + \vec f_R))_z - \kappa(\theta - \phi)  \right]  \ , \\
	\ddot\phi &= \frac{1}{i_2} \left[ \kappa(\theta - \phi) + (\vec r_R \times \vec f_R)_z + (\vec r_L \times \vec f_L)_z \right]  \ .
\end{align}
\ese
The subscript $z$ on a vector denotes the $z$--component. 

The equations of motion for the lower asperite are identical to those in Eqs.~(\ref{accelerations}), 
apart from the use of capital letters and the equilibrium position for the torsion spring. 
In particular,  the terms $\kappa(\theta - \phi)$ in Eqs.~(\ref{accelerations})  are replaced with $K(\Theta - \Phi  - \pi)$.

\section{Work and energy}\label{Sec:work}
The power delivered to the upper asperite  by friction  is 
$p_1(t) = \vec f_1 \cdot \dot{\vec r}_1$, 
where $\dot{\vec r}_1 = \dot{\vec r}_{\rm cm} + \ell_1 \dot\theta \langle \cos\theta, \sin\theta, 0 \rangle$ is the velocity of the upper pendulum bob.  The accumulated work  done by friction, as a function of time, is the integrated power:
\be
	w_1(t) =  \int_0^t dt\, p_1(t)  \ .
\ee
In turn,  power is the derivative of work: $p_1(t) = dw_1(t)/dt$. 
Analogous 
expressions hold for the power and work done by the applied forces $\vec f_2$, $\vec f_L$ and $\vec f_R$. 

The energy of the system (the upper asperite) can be split into  translational kinetic energy of the center of mass, 
\be
	k_{\rm trans} = \frac{1}{2}(m_1 + m_2)  \dot{\vec r}_{\rm cm}  \cdot \dot{\vec r}_{\rm cm} \ ,
\ee
and internal energy 
\be
	e_{\rm int} =  \frac{1}{2} (m_1 \ell_1^2 + m_2 \ell_2^2)  \dot\theta^2 + \frac{1}{2} i_2 \dot\phi^2 + \frac{1}{2} \kappa (\theta - \phi)^2 \ .
\ee
The first term in $e_{\rm int}$ is the rotational kinetic energy of the asperite about its center of mass with the bar treated as a point particle at $\vec r_2$. 
The second term is the rotational kinetic energy of the 
bar about its own center of mass. The third term is the potential energy stored in the torsion spring. 

The  net power is equal to the time rate of change of energy: 
\be\label{powerdEdt}
	p_1 + p_2 + p_L + p_R = \frac{d}{dt} ( k_{\rm trans} + e_{\rm int}) \ .
\ee
One can verify that Eq.~(\ref{powerdEdt}) holds when the equations of motion (\ref{accelerations}) are satisfied. 

The integral form of Eq.~(\ref{powerdEdt}) is the {\em work--energy principle}, 
\be\label{wetheorem}
	 w   = \Delta ( k_{\rm trans} + e_{\rm int})  \ .
\ee
The left--hand side is the net work done by external forces, $w = w_1 + w_2 + w_L + w_R$, and
the right--hand side is the change in total energy. 
Equation (\ref{wetheorem}) is the first law of thermodynamics applied to our purely mechanical model. For more  general systems, 
the first law of thermodynamics includes thermal energy, field energy, chemical energy, {\em etc.} 

The pseudopower delivered by the force of friction $\vec f_1$ is defined by $p_{1,\rm ps}(t) = \vec f_1 \cdot \dot{\vec r}_{\rm cm}$. 
This is the power that would be delivered to the system 
if $\vec f_1$ acted at the system's center of mass. Pseudowork \cite{Erlichson,Penchina,Sherwood}   is the time integral of pseudopower,
\be	
	w_{1,\rm ps}(t) = \int_0^t dt \, p_{1,\rm ps}(t)  \ .
\ee 
($w_{2,\rm ps}$, $w_{L,\rm ps}$ and $w_{R,\rm ps}$ are defined analogously.)
Multiplying Newton's second law Eq.~(\ref{NSL}) by $\dot{\vec r}_{\rm cm}$,  we find 
$p_{1,\rm ps}  + p_{2,\rm ps}  + p_{L,\rm ps}  + p_{R,\rm ps}  = d k_{\rm trans}/dt $. Integration gives 
the {\em pseudowork--energy principle},\cite{Erlichson,Penchina,Sherwood}  
\be\label{psworkenergythm}
	w_{\rm ps} = \Delta k_{\rm trans}   \ ,
\ee
where $w_{\rm ps} = w_{1,\rm ps} + w_{2,\rm ps}  + w_{L,\rm ps}  + w_{R,\rm ps}$ is the net pseudowork.   

By combining the work--energy principle and the pseudowork--energy principle, we see that the difference between  real work and  pseudowork
equals the change in internal energy, 
\be\label{Dewminuswps}
	w - w_{\rm ps} = \Delta e_{\rm int}  \ .
\ee
Sherwood and Bernard\cite{SherwoodBernard} refer to this result as the {\em invariant first law of thermodynamics} (IFLT). 
Roughly speaking, the IFLT says that the change in internal energy of a system is determined by the part of the work that is not responsible 
for changing the system's translational kinetic energy. 


\section{Forces}\label{Sec:forces}
For the  simulations presented in this paper, we choose 
\be\label{fLfR}
	\vec f_L = -\vec f_R  =  \frac{\kappa(\theta - \phi)}{a} \langle  -\sin\phi , \cos\phi, 0 \rangle  \ .
\ee
These forces act on the ends of the bar and insure that the net torque on the bar is zero. 
Equation (\ref{accelerations}c) shows  that  the angular acceleration of the bar vanishes, $\ddot\phi  = 0$.  
Since the initial angle and initial angular velocity of the bar are set to zero, the bar maintains its 
orientation $\phi(t) =0$ throughout each simulation.  

For forced sliding, we choose $\vec f_2 = - \vec f_1$. In this case the net force 
$\vec f_1 + \vec f_2 + \vec f_L + \vec f_R$ on the upper asperite vanishes and Newton's second law (\ref{NSL})  yields $\ddot{\vec r}_{\rm cm} = 0$. 
Analogous relations hold for the forces on the lower asperite. Thus, $\Phi(t) = 0$  
and $\ddot{\vec R}_{\rm cm} = 0$. 

With free sliding, the net force on the lower asperite is again zero. 
For the upper asperite, $\vec f_2$ is modified to  
$\vec f_2 = \langle 0,  -f_{1,y},0 \rangle$. Thus, $\vec f_1$ and $\vec f_2$ are balanced in the 
$y$ and $z$--directions, but  not in the $x$--direction. The upper asperite is allowed to 
slow down or speed up due to the unbalanced component $f_{1,x}$ of the friction force. 

The forces $\vec f_1$ and $\vec F_1$ are determined by the interaction between the pendulum bobs on the upper and lower asperites. 
For an electric interaction, the potential is 
\be\label{Uelec}
	U_{\rm elec} = \frac{1}{4\pi\epsilon_0} \frac{qQ}{\mathcal{R}}
\ee
where $\mathcal{R}$ is  the magnitude of the separation vector 
$\vec{\mathcal{R}} \equiv \vec r_1 - \vec R_1$ between pendulum bobs and  $q$ and $Q$ are the charges on the bobs. The electric forces are 
repulsive when $qQ>0$, and attractive when $qQ<0$. 

\section{Numerical Implementation}\label{Sec:numerics}
The equations of motion (\ref{accelerations}) are converted to first order by defining the velocities
${\vec v}_{\rm cm} \equiv \dot{\vec r}_{\rm cm}$,  $\omega \equiv \dot\theta$ and $\nu \equiv \dot\phi$ as independent variables. Integration  
is carried out  using a simple second--order Runge--Kutta scheme.  For each force,  power is computed from the half-timestep 
positions and velocities. The work in each timestep is computed   as the product of power and the timestep interval $\Delta t$. 
The results from this numerical scheme are second--order convergent. That is, the errors are 
proportional to $\Delta t^2$. 


\section{Simulation Results}\label{Sec:ExamplesResults}
\subsection{Forced sliding with repulsive interaction}\label{Sec:ForcedElecRepel}
We choose 
parameter values for the asperites  somewhat arbitrarily. In this section,  
$m_1 = M_1 = 1$, $m_2 = M_2 = 2$, $\kappa = K =  2$, $\ell = L = 2$, $a = A = 3$, and $i_2= I_2 = 5$.  (Units are unspecified.) 
The interaction   is a repulsive electric force with $qQ/(4\pi\epsilon_0) = 1/5$. 

The asperites are given initial positions  $\vec r_{\rm cm} = \langle -10,\ell_1,0 \rangle$, $\vec R_{\rm cm} = \langle 0, -L_1,0 \rangle$, and
 initial velocities  $\vec v_{\rm cm} = \langle 1/2, 0, 0 \rangle$,  $\vec V_{\rm cm} = \langle 0, 0, 0 \rangle$. The initial values for the
 angles are $\theta = 0$, $\phi = 0$,
 $\Theta = \pi$, and $\Phi = 0$, and the initial angular velocities are all zero. 
  
\begin{figure}[h]
\centering
\includegraphics[scale=1]{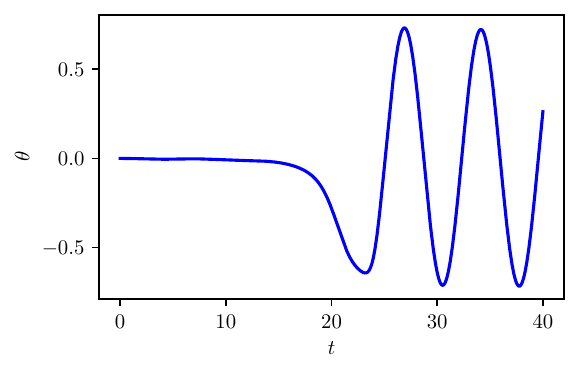}
\caption{Pendulum angle $\theta$ as a function of time $t$ for the upper asperite. The interaction between asperites is electric and repulsive.}
\label{thetaelecfig}
\end{figure}

\begin{figure}[h!]
\begin{tikzpicture}[scale=0.6]
	\begin{scope}[shift={(0,0)}]
	\def\a{3}; 
	\def\lone{1.333};
	\def\ltwo{0.667};
	\def\th{-0.28065*180/3.14159};
	\coordinate (rcm) at (0,0); 
	\coordinate (r1) at ($(rcm) + ({\lone*sin(\th)}, {-\lone*cos(\th)})$);
	\coordinate (r2) at ($(rcm) + ({-\ltwo*sin(\th)},{\ltwo*cos(\th)})$);
	\draw[thick, fill=blue!20] ($(r2) + (-\a/2, - 0.1)$) rectangle ($(r2) + (\a/2, 0.1)$);   
	\draw[thick, rotate=\th, fill=yellow!20] ($(rcm) + (0.05,-\lone)$) rectangle  ($(rcm) + (-0.05,\ltwo)$);  
	\draw[thick, fill=red!40] (rcm) circle (0.07);           
	\draw [thick, fill=blue!20] (r1) circle (0.18);                                                    
	\draw[fill] (r2) circle (0.05);                                                                                     
	\end{scope}
	\begin{scope}[shift={(0,-2.667)}]
	\def\A{3};
	\def\Lone{1.333};
	\def\Ltwo{0.667};
	\def\Th{2.8609*180/3.14159};
	\coordinate (Rcm) at (0,0); 
	\coordinate (R1) at ($(Rcm) + ({\Lone*sin(\Th)}, {-\Lone*cos(\Th)})$);
	\coordinate (R2) at ($(Rcm) + ({-\Ltwo*sin(\Th)},{\Ltwo*cos(\Th)})$);
	\draw[thick, fill=blue!20] ($(R2) + (-\A/2, - 0.1)$) rectangle ($(R2) + (\A/2, 0.1)$);   
	\draw[thick, rotate=\Th, fill=yellow!20] ($(Rcm) + (0.05,-\Lone)$) rectangle  ($(Rcm) + (-0.05,\Ltwo)$);  
	\draw[thick, fill=red!40] (Rcm) circle (0.07);           
	\draw[dashed] (0,-1.2) -- (0,3.8); 
	\draw [thick, fill=blue!20] (R1) circle (0.18);                                                    
	\draw[fill] (R2) circle (0.05);               
	\end{scope}
\end{tikzpicture}
\quad
\begin{tikzpicture}[scale=0.6]
	\begin{scope}[shift={(0.75,0)}]
	\def\a{3}; 
	\def\lone{1.333};
	\def\ltwo{0.667};
	\def\th{-0.52731*180/3.14159};
	\coordinate (rcm) at (0,0); 
	\coordinate (r1) at ($(rcm) + ({\lone*sin(\th)}, {-\lone*cos(\th)})$);
	\coordinate (r2) at ($(rcm) + ({-\ltwo*sin(\th)},{\ltwo*cos(\th)})$);
	\draw[thick, fill=blue!20] ($(r2) + (-\a/2, - 0.1)$) rectangle ($(r2) + (\a/2, 0.1)$);   
	\draw[thick, rotate=\th, fill=yellow!20] ($(rcm) + (0.05,-\lone)$) rectangle  ($(rcm) + (-0.05,\ltwo)$);  
	\draw[thick, fill=red!40] (rcm) circle (0.07);           
	\draw [thick, fill=blue!20] (r1) circle (0.18);                                                    
	\draw[fill] (r2) circle (0.05);                                                                                     
	\end{scope}
	\begin{scope}[shift={(0,-2.667)}]
	\def\A{3};
	\def\Lone{1.333};
	\def\Ltwo{0.667};
	\def\Th{2.6143*180/3.14159};
	\coordinate (Rcm) at (0,0); 
	\coordinate (R1) at ($(Rcm) + ({\Lone*sin(\Th)}, {-\Lone*cos(\Th)})$);
	\coordinate (R2) at ($(Rcm) + ({-\Ltwo*sin(\Th)},{\Ltwo*cos(\Th)})$);
	\draw[thick, fill=blue!20] ($(R2) + (-\A/2, - 0.1)$) rectangle ($(R2) + (\A/2, 0.1)$);   
	\draw[thick, rotate=\Th, fill=yellow!20] ($(Rcm) + (0.05,-\Lone)$) rectangle  ($(Rcm) + (-0.05,\Ltwo)$);  
	\draw[thick, fill=red!40] (Rcm) circle (0.07);           
	\draw[dashed] (0,-1.2) -- (0,3.8); 
	\draw [thick, fill=blue!20] (R1) circle (0.18);                                                    
	\draw[fill] (R2) circle (0.05);               
	\end{scope}
\end{tikzpicture}
\begin{tikzpicture}[scale=0.6]
	\begin{scope}[shift={(1.50,0)}]
	\def\a{3}; 
	\def\lone{1.333};
	\def\ltwo{0.667};
	\def\th{-0.64080*180/3.14159};
	\coordinate (rcm) at (0,0); 
	\coordinate (r1) at ($(rcm) + ({\lone*sin(\th)}, {-\lone*cos(\th)})$);
	\coordinate (r2) at ($(rcm) + ({-\ltwo*sin(\th)},{\ltwo*cos(\th)})$);
	\draw[thick, fill=blue!20] ($(r2) + (-\a/2, - 0.1)$) rectangle ($(r2) + (\a/2, 0.1)$);   
	\draw[thick, rotate=\th, fill=yellow!20] ($(rcm) + (0.05,-\lone)$) rectangle  ($(rcm) + (-0.05,\ltwo)$);  
	\draw[thick, fill=red!40] (rcm) circle (0.07);           
	\draw [thick, fill=blue!20] (r1) circle (0.18);                                                    
	\draw[fill] (r2) circle (0.05);                                                                                     
	\end{scope}
	\begin{scope}[shift={(0,-2.667)}]
	\def\A{3};
	\def\Lone{1.333};
	\def\Ltwo{0.667};
	\def\Th{2.50079*180/3.14159};
	\coordinate (Rcm) at (0,0); 
	\coordinate (R1) at ($(Rcm) + ({\Lone*sin(\Th)}, {-\Lone*cos(\Th)})$);
	\coordinate (R2) at ($(Rcm) + ({-\Ltwo*sin(\Th)},{\Ltwo*cos(\Th)})$);
	\draw[thick, fill=blue!20] ($(R2) + (-\A/2, - 0.1)$) rectangle ($(R2) + (\A/2, 0.1)$);   
	\draw[thick, rotate=\Th, fill=yellow!20] ($(Rcm) + (0.05,-\Lone)$) rectangle  ($(Rcm) + (-0.05,\Ltwo)$);  
	\draw[thick, fill=red!40] (Rcm) circle (0.07);           
	\draw[dashed] (0,-1.2) -- (0,3.8); 
	\draw [thick, fill=blue!20] (R1) circle (0.18);                                                    
	\draw[fill] (R2) circle (0.05);               
	\end{scope}
\end{tikzpicture}
\begin{tikzpicture}[scale=0.6]
	\begin{scope}[shift={(2.25,0)}]
	\def\a{3}; 
	\def\lone{1.333};
	\def\ltwo{0.667};
	\def\th{-0.33425*180/3.14159};
	\coordinate (rcm) at (0,0); 
	\coordinate (r1) at ($(rcm) + ({\lone*sin(\th)}, {-\lone*cos(\th)})$);
	\coordinate (r2) at ($(rcm) + ({-\ltwo*sin(\th)},{\ltwo*cos(\th)})$);
	\draw[thick, fill=blue!20] ($(r2) + (-\a/2, - 0.1)$) rectangle ($(r2) + (\a/2, 0.1)$);   
	\draw[thick, rotate=\th, fill=yellow!20] ($(rcm) + (0.05,-\lone)$) rectangle  ($(rcm) + (-0.05,\ltwo)$);  
	\draw[thick, fill=red!40] (rcm) circle (0.07);           
	\draw [thick, fill=blue!20] (r1) circle (0.18);                                                    
	\draw[fill] (r2) circle (0.05);                                                                                     
	\end{scope}
	\begin{scope}[shift={(0,-2.667)}]
	\def\A{3};
	\def\Lone{1.333};
	\def\Ltwo{0.667};
	\def\Th{2.8073*180/3.14159};
	\coordinate (Rcm) at (0,0); 
	\coordinate (R1) at ($(Rcm) + ({\Lone*sin(\Th)}, {-\Lone*cos(\Th)})$);
	\coordinate (R2) at ($(Rcm) + ({-\Ltwo*sin(\Th)},{\Ltwo*cos(\Th)})$);
	\draw[thick, fill=blue!20] ($(R2) + (-\A/2, - 0.1)$) rectangle ($(R2) + (\A/2, 0.1)$);   
	\draw[thick, rotate=\Th, fill=yellow!20] ($(Rcm) + (0.05,-\Lone)$) rectangle  ($(Rcm) + (-0.05,\Ltwo)$);  
	\draw[thick, fill=red!40] (Rcm) circle (0.07);           
	\draw[dashed] (0,-1.2) -- (0,3.8); 
	\draw [thick, fill=blue!20] (R1) circle (0.18);                                                    
	\draw[fill] (R2) circle (0.05);               
	\end{scope}
\end{tikzpicture}
\quad
\begin{tikzpicture}[scale=0.6]
	\begin{scope}[shift={(3.00,0)}]
	\def\a{3}; 
	\def\lone{1.333};
	\def\ltwo{0.667};
	\def\th{0.52383*180/3.14159};
	\coordinate (rcm) at (0,0); 
	\coordinate (r1) at ($(rcm) + ({\lone*sin(\th)}, {-\lone*cos(\th)})$);
	\coordinate (r2) at ($(rcm) + ({-\ltwo*sin(\th)},{\ltwo*cos(\th)})$);
	\draw[thick, fill=blue!20] ($(r2) + (-\a/2, - 0.1)$) rectangle ($(r2) + (\a/2, 0.1)$);   
	\draw[thick, rotate=\th, fill=yellow!20] ($(rcm) + (0.05,-\lone)$) rectangle  ($(rcm) + (-0.05,\ltwo)$);  
	\draw[thick, fill=red!40] (rcm) circle (0.07);           
	\draw [thick, fill=blue!20] (r1) circle (0.18);                                                    
	\draw[fill] (r2) circle (0.05);                                                                                     
	\end{scope}
	\begin{scope}[shift={(0,-2.667)}]
	\def\A{3};
	\def\Lone{1.333};
	\def\Ltwo{0.667};
	\def\Th{3.6654*180/3.14159};
	\coordinate (Rcm) at (0,0); 
	\coordinate (R1) at ($(Rcm) + ({\Lone*sin(\Th)}, {-\Lone*cos(\Th)})$);
	\coordinate (R2) at ($(Rcm) + ({-\Ltwo*sin(\Th)},{\Ltwo*cos(\Th)})$);
	\draw[thick, fill=blue!20] ($(R2) + (-\A/2, - 0.1)$) rectangle ($(R2) + (\A/2, 0.1)$);   
	\draw[thick, rotate=\Th, fill=yellow!20] ($(Rcm) + (0.05,-\Lone)$) rectangle  ($(Rcm) + (-0.05,\Ltwo)$);  
	\draw[thick, fill=red!40] (Rcm) circle (0.07);           
	\draw[dashed] (0,-1.2) -- (0,3.8); 
	\draw [thick, fill=blue!20] (R1) circle (0.18);                                                    
	\draw[fill] (R2) circle (0.05);               
	\end{scope}
\end{tikzpicture}
\caption{The model asperites at times $t = 20.0$, $21.5$, $23.0$, $24.5$ and $26.0$ for the repulsive electric interaction. The dashed lines pass 
through the center of mass of the lower asperite, which remains in place throughout the simulation.}
\label{Elec20thru26}
\end{figure}
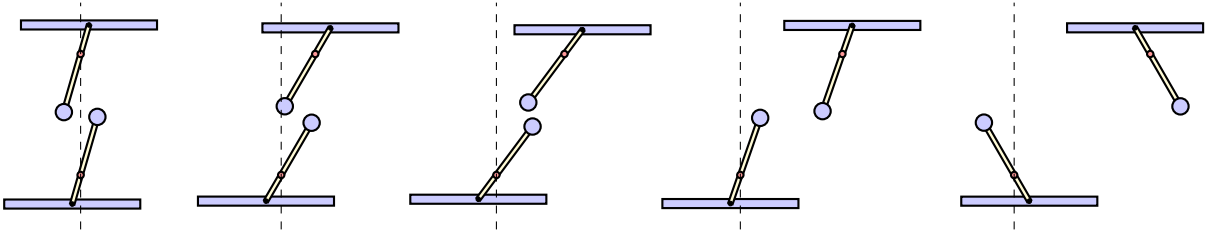

Figure \ref{thetaelecfig} shows the pendulum angle $\theta$ for the upper asperite  as a function of time.  
The centers of mass for the two asperites pass one another at time $t=20$. 
Figure \ref{Elec20thru26} shows a sequence of snapshots at equally spaced times 
from $t=20$ through $t=26$. The interaction sets the 
asperites into motion, oscillating about their equilibrium positions. 
The internal energy of each asperite  is a combination 
of kinetic energy of the oscillating pendulum and potential energy in the stretched/compressed torsion spring. If each asperite was part of a macroscopic 
object (such as a block or a table), the internal energy  would dissipate into thermal energy. 

\begin{figure}[htb]
\centering
\includegraphics[scale=1]{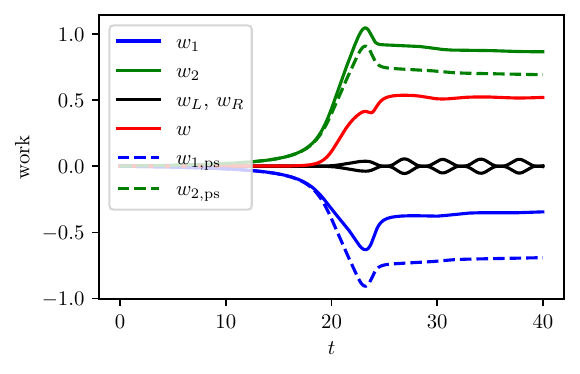}
\caption{The work done on the upper asperite for 
forced sliding with repulsive electric interaction.  The curves are: work by friction ($w_1$); work by applied forces ($w_2$, $w_L$ and $w_R$); net 
work ($w$); pseudowork by friction ($w_{1,{\rm ps}}$); and pseudowork by  $\vec f_2$ ($w_{2,{\rm ps}})$. }
\label{workpowerrepulsionfig}
\end{figure}

Figure \ref{workpowerrepulsionfig} shows the  work done on the upper asperite as a function of time by each of the external forces. 
The work done by friction is negative, $w_1 < 0$,  whereas the work done by the applied force
 $\vec f_2$ is positive, $w_2 > 0$. The work 
done by the forces $\vec f_L$ and $\vec f_R$  cancel,  $w_L + w_R = 0$. 

 The net work  on  the upper asperite is $w = 0.520$ and the change in the asperite's internal energy is $\Delta e_{\rm int} = 0.520$. 
 For forced sliding, $\Delta k_{\rm trans} = 0$ . 
Thus, $w$ agrees (to within numerical error) with 
the change in total energy $\Delta(k_{\rm trans} + e_{\rm int})$, confirming the work--energy principle (\ref{wetheorem}).  

With forced sliding the applied force $\vec f_2$ is equal in magnitude and opposite in direction to the friction force $\vec f_1$. 
Also, $\vec f_L + \vec f_R = 0$. Therefore the net {\em impulse} vanishes, and the center of mass of the upper asperite maintains a constant 
velocity throughout the interaction. The net {\em work}, however,  is not zero because the distances through which the forces 
$\vec f_1$ and $\vec f_2$ act are different. The force $\vec f_2$ acts 
through the distance that the bar moves,  whereas $\vec f_1$ acts through the distance that the pendulum bob moves. 

Ultimately we would like to understand why the internal energy of the upper asperite increases due to its interaction with the lower asperite. 
For the internal energy to increase,  the IFLT Eq.~(\ref{Dewminuswps}) tells us that the net work must be greater than the net pseudowork. 
The following analysis shows 
why.   

The pendulum bobs pass one another at $t \approx 23.2$, when $w_1$ reaches a minimum (the power $p_1$ switches from negative to positive)
and $w_2$ reaches a maximum ($p_2$ switches from positive to negative). 
This time marks the separation between an 
``approach phase," when the pendulum bobs are approaching
one another, and a ``recede phase," when the bobs are receding from one another. 

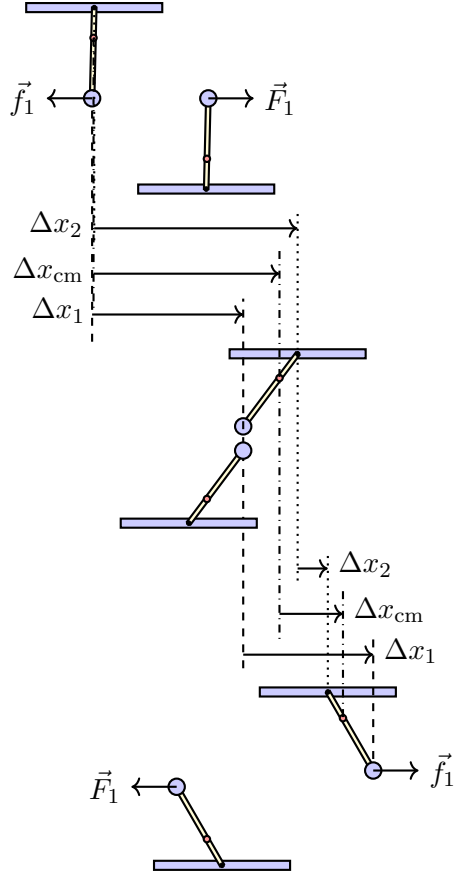
\begin{figure}[htb]
\begin{tikzpicture}[scale=0.6]
	\begin{scope}[shift={(-2.5,0)}]
	\def\a{3}; 
	\def\lone{1.333};
	\def\ltwo{0.667};
	\def\th{-0.02340*180/3.14159};
	\coordinate (rcm) at (0,0); 
	\coordinate (r1) at ($(rcm) + ({\lone*sin(\th)}, {-\lone*cos(\th)})$);
	\coordinate (r2) at ($(rcm) + ({-\ltwo*sin(\th)},{\ltwo*cos(\th)})$);
	\draw[thick, fill=blue!20] ($(r2) + (-\a/2, - 0.1)$) rectangle ($(r2) + (\a/2, 0.1)$);   
	\draw[thick, rotate=\th, fill=yellow!20] ($(rcm) + (0.05,-\lone)$) rectangle  ($(rcm) + (-0.05,\ltwo)$);  
	\draw[thick, fill=red!40] (rcm) circle (0.07);           
	\draw [thick, fill=blue!20] (r1) circle (0.18);                                                    
	\draw[->,thick] (r1) -- ($(r1) - (1,0)$) node[left] {$\vec f_1$}; 
	\draw[fill] (r2) circle (0.05);                                                                                     
	\draw [dashed,thick] ($(rcm) + ({\lone*sin(\th)}, {-6.7})$) -- ($(rcm) + ({\lone*sin(\th)}, {-1.5})$);  
	\draw[->,thick] ($({\lone*sin(\th)}, {-6.1})$)  --+ (3.3314,0) node[left, shift={(-1.95,0.05)}] {$\Delta x_1$};     
	\draw[dotted,thick] (r2) -- ($(r2) + (0,-5.25)$);   
	\draw[->,thick] ($({-\ltwo*sin(\th)},{-4.2})$)  --+ (4.4844,0) node[left, shift={(-2.7,0.05)}] {$\Delta x_2$};  
	\draw[dashdotted,thick] (rcm) -- ($(rcm) + (0,-6.0)$);   
	\draw[->,thick] ($(0,-5.2)$)  --+ (4.1,0) node[left, shift={(-2.45,0.05)}] {$\Delta x_{\rm cm}$};  
	\end{scope}
	\begin{scope}[shift={(0,-2.667)}]
	\def\A{3};
	\def\Lone{1.333};
	\def\Ltwo{0.667};
	\def\Th{3.1182*180/3.14159};
	\coordinate (Rcm) at (0,0); 
	\coordinate (R1) at ($(Rcm) + ({\Lone*sin(\Th)}, {-\Lone*cos(\Th)})$);
	\coordinate (R2) at ($(Rcm) + ({-\Ltwo*sin(\Th)},{\Ltwo*cos(\Th)})$);
	\draw[thick, fill=blue!20] ($(R2) + (-\A/2, - 0.1)$) rectangle ($(R2) + (\A/2, 0.1)$);   
	\draw[thick, rotate=\Th, fill=yellow!20] ($(Rcm) + (0.05,-\Lone)$) rectangle  ($(Rcm) + (-0.05,\Ltwo)$);  
	\draw[thick, fill=red!40] (Rcm) circle (0.07);           
	\draw [thick, fill=blue!20] (R1) circle (0.18);                                                    
	\draw[->,thick] (R1) -- ($(R1) + (1,0)$) node[right] {$\vec F_1$}; 
	\draw[fill] (R2) circle (0.05);               
	\end{scope}
	\begin{scope}[shift={(1.600,-7.5)}]
	\def\a{3}; 
	\def\lone{1.333};
	\def\ltwo{0.667};
	\def\th{-0.6435*180/3.14159};
	\coordinate (rcm) at (0,0); 
	\coordinate (r1) at ($(rcm) + ({\lone*sin(\th)}, {-\lone*cos(\th)})$);
	\coordinate (r2) at ($(rcm) + ({-\ltwo*sin(\th)},{\ltwo*cos(\th)})$);
	\draw[thick, fill=blue!20] ($(r2) + (-\a/2, - 0.1)$) rectangle ($(r2) + (\a/2, 0.1)$);   
	\draw[thick, rotate=\th, fill=yellow!20] ($(rcm) + (0.05,-\lone)$) rectangle  ($(rcm) + (-0.05,\ltwo)$);  
	\draw[thick, fill=red!40] (rcm) circle (0.07);           
	\draw [thick, fill=blue!20] (r1) circle (0.18);                                                    

	\draw[fill] (r2) circle (0.05);                                                                                     
	\draw [dashed,thick] ($(rcm) + ({\lone*sin(\th)}, {-6.4})$) -- ($(rcm) + ({\lone*sin(\th)}, {1.75})$);  
	\draw[->,thick] ($({\lone*sin(\th)}, {-6.1})$)  --+ (2.8694,0) node[right, shift={(0.0,0.05)}] {$\Delta x_1$};     
	\draw[dotted,thick] ($(r2) + (0,-5)$) -- ($(r2) + (0,3.1)$);    
	\draw[->,thick] ($({-\ltwo*sin(\th)},{-4.2})$)  --+ (0.6653,0) node[right, shift={(0,0.05)}] {$\Delta x_2$};  
	\draw[dashdotted,thick] ($(rcm) + (0,2.8)$) -- ($(rcm) + (0,-5.8)$);   
	\draw[->,thick] ($(0,-5.2)$)  --+ (1.4,0) node[right, shift={(0.0,0.05)}] {$\Delta x_{\rm cm}$};  
	\end{scope}
	\begin{scope}[shift={(0,-2.667-7.5)}]
	\def\A{3};
	\def\Lone{1.333};
	\def\Ltwo{0.667};
	\def\Th{2.4981*180/3.14159};
	\coordinate (Rcm) at (0,0); 
	\coordinate (R1) at ($(Rcm) + ({\Lone*sin(\Th)}, {-\Lone*cos(\Th)})$);
	\coordinate (R2) at ($(Rcm) + ({-\Ltwo*sin(\Th)},{\Ltwo*cos(\Th)})$);
	\draw[thick, fill=blue!20] ($(R2) + (-\A/2, - 0.1)$) rectangle ($(R2) + (\A/2, 0.1)$);   
	\draw[thick, rotate=\Th, fill=yellow!20] ($(Rcm) + (0.05,-\Lone)$) rectangle  ($(Rcm) + (-0.05,\Ltwo)$);  
	\draw[thick, fill=red!40] (Rcm) circle (0.07);           
	\draw [thick, fill=blue!20] (R1) circle (0.18);                                                    
	\draw[fill] (R2) circle (0.05);               
	\end{scope}
	\begin{scope}[shift={(3.0,-15)}]
	\def\a{3}; 
	\def\lone{1.333};
	\def\ltwo{0.667};
	\def\th{0.5238*180/3.14159};
	\coordinate (rcm) at (0,0); 
	\coordinate (r1) at ($(rcm) + ({\lone*sin(\th)}, {-\lone*cos(\th)})$);
	\coordinate (r2) at ($(rcm) + ({-\ltwo*sin(\th)},{\ltwo*cos(\th)})$);
	\draw[thick, fill=blue!20] ($(r2) + (-\a/2, - 0.1)$) rectangle ($(r2) + (\a/2, 0.1)$);   
	\draw[thick, rotate=\th, fill=yellow!20] ($(rcm) + (0.05,-\lone)$) rectangle  ($(rcm) + (-0.05,\ltwo)$);  
	\draw[thick, fill=red!40] (rcm) circle (0.07);           
	\draw [thick, fill=blue!20] (r1) circle (0.18);                                                    
	\draw[->,thick] (r1) -- ($(r1) + (1,0)$) node[right] {$\vec f_1$}; 
	\draw[fill] (r2) circle (0.05);                                                                                     
	\draw [dashed,thick] ($(rcm) + ({\lone*sin(\th)}, {-0.9})$) -- ($(rcm) + ({\lone*sin(\th)}, {1.8})$);  
	\draw[dotted,thick] ($(r2) + (0,0)$) -- ($(r2) + (0,3)$);    
	\draw[dashdotted,thick] ($(rcm) + (0,2.8)$) -- ($(rcm) + (0,0)$);   
	\end{scope}
	\begin{scope}[shift={(0,-2.667-15)}]
	\def\A{3};
	\def\Lone{1.333};
	\def\Ltwo{0.667};
	\def\Th{3.6654*180/3.14159};
	\coordinate (Rcm) at (0,0); 
	\coordinate (R1) at ($(Rcm) + ({\Lone*sin(\Th)}, {-\Lone*cos(\Th)})$);
	\coordinate (R2) at ($(Rcm) + ({-\Ltwo*sin(\Th)},{\Ltwo*cos(\Th)})$);
	\draw[thick, fill=blue!20] ($(R2) + (-\A/2, - 0.1)$) rectangle ($(R2) + (\A/2, 0.1)$);   
	\draw[thick, rotate=\Th, fill=yellow!20] ($(Rcm) + (0.05,-\Lone)$) rectangle  ($(Rcm) + (-0.05,\Ltwo)$);  
	\draw[thick, fill=red!40] (Rcm) circle (0.07);           
	\draw [thick, fill=blue!20] (R1) circle (0.18);                                                    
	\draw[->,thick] (R1) -- ($(R1) - (1,0)$) node[left] {$\vec F_1$}; 
	\draw[fill] (R2) circle (0.05);               
	\end{scope}
\end{tikzpicture}
\caption{Forced sliding with repulsive electric interaction. The sequence shows the asperites at times $t = 15.0$ (top figure), $23.2$ (middle 
figure) and $26.0$ (bottom figure). Displacements for the upper asperite are shown for the approach phase (from top to middle) and for the 
recede phase (from middle to bottom). }
\label{ForcedRepulsiveFig}
\end{figure}

Now observe: The work done on the asperite remains close to zero until, very roughly, $t=15$.  Beyond about $t=26$, the work remains largely unchanged. 
Let us define the approach phase by $15.0 \le t \le 23.2$ and the recede phase by $23.2 \le t \le 26.0$. 
Figure \ref{ForcedRepulsiveFig}  shows the asperites at  times $t = 15.0$, $23.2$ and $26.0$.  The figure includes labels for the 
$x$--components of the displacements of the pendulum bob $\Delta x_1$, bar $\Delta x_2$, and center of mass $\Delta x_{\rm cm}$  for both phases. 

Consider first the approach phase. 
The center of mass of the upper asperite moves a distance $\Delta x_{\rm cm} = 4.10$. 
Because the asperite is not rigid, the repulsive electric force causes the 
pendulum bob to lag behind the center of mass. Thus, during this phase, 
the bob moves a relatively small distance $\Delta x_1 \approx 3.33$. Since the friction force $\vec f_1$ is 
primarily in the $-x$ direction, both the work $w_1 \approx f_{1,x} \Delta x_1$ and pseudowork 
$w_{1,{\rm ps}} \approx f_{1,x} \Delta x_{\rm cm}$ are negative. 
From $\Delta x_{\rm cm} > \Delta x_1$ it follows that $w_1 > w_{1,{\rm ps}}$. 

Also during the approach phase,  the applied force $\vec f_2$ pushes the bar  ahead of the center of mass. 
The bar moves a relatively large distance $\Delta x_2 \approx 4.48$, so $\Delta x_2 > \Delta x_{\rm cm}$. Since $\vec f_2$ is primarily in the 
$+x$ direction, both the work $w_2 \approx f_{2,x} \Delta x_2$ and pseudowork $w_{2,{\rm ps}} \approx f_{2,x} \Delta x_{\rm cm}$ 
are positive. It follows that $w_2 > w_{2,{\rm ps}}$. 

Now consider the recede phase. The center of mass displacement  shown in Fig.~\ref{ForcedRepulsiveFig} is  
$\Delta x_{\rm cm} = 1.40$. The pendulum bob 
moves a larger distance  $\Delta x_1 \approx 2.87$ due to the repulsive friction force. The work and pseudowork are positive 
so that $w_1 > w_{1,{\rm ps}}$. 
During this same phase the bar moves a relatively small distance $\Delta x_2 \approx 0.670$ due to the applied force $\vec f_2$. 
The work and pseudowork are both negative, so that   $w_2 > w_{2,{\rm ps}}$. 

Through both the approach and recede phases of motion, the work done by each force $\vec f_1$ and $\vec f_2$ is greater 
than the corresponding pseudowork. 
 In Fig.~\ref{workpowerrepulsionfig}  the pseudowork  
curves (dashed) lie below the corresponding work curves (solid). 

It is worth noting that with forced sliding the net pseudowork vanishes, $w_{\rm ps} = w_{1,{\rm ps}} + w_{2,{\rm ps}} =  \vec f_1 \cdot \Delta \vec r_{\rm cm}
+ \vec f_2 \cdot \Delta r_{\rm cm} = 0$, because $\vec f_2 = -\vec f_1$. Therefore  the net work must be positive, $w > 0$. 
But the key point is that the net work is greater than the net pseudowork, and by the 
IFLT the difference $w - w_{\rm ps}$ equals the increase in internal energy of the asperite. As part of a physical block, the internal energy of the asperite dissipates into thermal energy. 

The analysis above is not rigorous. It suggests that the power delivered to the upper asperite is always positive. (That is, the slope of the 
curve $w(t)$ is always positive.) 
This is not precisely correct because the forces and displacements are not 
simply one--dimensional;  work also depends on the motion and forces in the $y$--direction. 
Moreover, inertial effects can conspire to reverse the inequality between work and pseudowork:  The displacement of the point of contact of a force, 
in the direction of the force,  is not 
always greater than the displacement of the object's center of mass. 
A close look at Figure \ref{workpowerrepulsionfig} shows that the net work actually  decreases (the power becomes negative) 
during brief periods of time; for example,  between $t = 23.2$ and $23.7$. 
Overall, however,  the work done by the end of the interaction is positive, and this is the source of increase in thermal energy. 

\subsection{Forced sliding with attractive interaction}\label{Sec:ForcedElecAttract}
In the previous subsection we saw that with repulsive forces the net work is greater than the net pseudowork  and the asperite's internal 
energy increases. 
One might guess that the work will reverse sign if the forces reverse sign.  This is not the case. 

In this subsection we consider attractive electric forces with $qQ/(4\pi\epsilon_0) = -1/5$. 
In addition, the masses and spring constants are modified to 
$m_1 = M_1 = 5$, $m_2 = M_2 = 3$ and $\kappa = K = 1$. These changes decrease the natural oscillation frequency of the torsion pendulums 
and prevent the pendulums from executing rapid oscillations during their interaction. This keeps the dynamics relatively simple. 

The initial conditions are the same as in subsec.~\ref{Sec:ForcedElecRepel} apart from the addition of a small
separation  $s = 0.3$ between the asperites  in the $y$--direction. That is, we  modify the initial position of the lower 
asperite  to $\vec R_{\rm cm} = \langle 0, -L_1 - s,0 \rangle$.   This separation allows us to 
avoid the singularity at $\mathcal{R} = 0$.   

\begin{figure}[htb]
\centering
\includegraphics[scale=1]{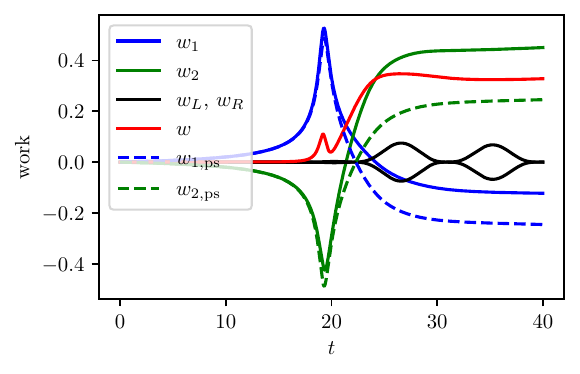}
\caption{The work done on the upper asperite for forced sliding with attractive electric interaction. }
\label{workpowerEAB}
\end{figure}

As expected, the interaction excites the asperites, causing their torsion pendulums to oscillate. 
Figure \ref{workpowerEAB} shows the work done on the upper asperite. 
At the end of the simulation, the net work is positive, $w = 0.347$. This agrees, to within numerical error, with the increase in total energy. 
(In particular,  $\Delta e_{\rm int} = 0.347$ and with forced sliding, $\Delta k_{\rm trans} = 0$.) The work--energy principle (\ref{wetheorem}) holds. 

Since the internal energy of the asperite increases, work must be greater than pseudowork. Here is why.  First, observe that 
the pendulum bobs pass one another at time $t \approx 19.3$, when $w_1$ reaches a maximum ($p_1$ switches from positive to negative) and $w_2$ reaches a minimum ($p_2$ switches from negative to positive). 
The work done on the upper asperite is negligible before about $t=15$, and after about $t=26$. We define 
the approach phase by $15.0  \le t  \le 19.3$ and the recede phase by $19.3 \le t  \le 26.0$. 

Figure \ref{ForcedRepulsiveFig} shows  the asperites   at times $t = 15.0$, $19.3$ and $26.0$. 
\begin{figure}[htb]
\begin{tikzpicture}[scale=0.6]
	\begin{scope}[shift={(-2.5,0)}]
	\def\a{3}; 
	\def\lone{0.75};
	\def\ltwo{1.25};
	\def\th{0.0333*180/3.14159};
	\coordinate (rcm) at (0,0); 
	\coordinate (r1) at ($(rcm) + ({\lone*sin(\th)}, {-\lone*cos(\th)})$);
	\coordinate (r2) at ($(rcm) + ({-\ltwo*sin(\th)},{\ltwo*cos(\th)})$);
	\draw[thick, fill=blue!20] ($(r2) + (-\a/2, - 0.1)$) rectangle ($(r2) + (\a/2, 0.1)$);   
	\draw[thick, rotate=\th, fill=yellow!20] ($(rcm) + (0.05,-\lone)$) rectangle  ($(rcm) + (-0.05,\ltwo)$);  
	\draw[thick, fill=red!40] (rcm) circle (0.07);           
	\draw [thick, fill=blue!20] (r1) circle (0.18);                                                    
	\draw[->,thick] (r1) -- ($(r1) + (0.9,-0.1)$) node[above] {$\vec f_1$}; 
	\draw[fill] (r2) circle (0.05);                                                                                     
	\draw [dashed,thick] ($(rcm) + ({\lone*sin(\th)}, {-6.4})$) -- ($(rcm) + ({\lone*sin(\th)}, {-1.0})$);  
	\draw[->,thick] ($({\lone*sin(\th)}, {-5.9})$)  --+ (2.300,0) node[left, shift={(-1.3,0.05)}] {$\Delta x_1$};     
	\draw[dotted,thick] (r2) -- ($(r2) + (0,-5.6)$);   
	\draw[->,thick] ($({-\ltwo*sin(\th)},{-3.9})$)  --+ (1.900,0) node[left, shift={(-1.1,0.05)}] {$\Delta x_2$};  
	\draw[dashdotted,thick] (rcm) -- ($(rcm) + (0,-5.6)$);   
	\draw[->,thick] (0,-4.9)  --+ (2.15,0) node[left, shift={(-1.25,0.05)}] {$\Delta x_{\rm cm}$};  
	\end{scope}
	\begin{scope}[shift={(0,-1.5-0.3)}]   
	\def\A{3};
	\def\Lone{0.75};
	\def\Ltwo{1.25};
	\def\Th{3.1749*180/3.14159};
	\coordinate (Rcm) at (0,0); 
	\coordinate (R1) at ($(Rcm) + ({\Lone*sin(\Th)}, {-\Lone*cos(\Th)})$);
	\coordinate (R2) at ($(Rcm) + ({-\Ltwo*sin(\Th)},{\Ltwo*cos(\Th)})$);
	\draw[thick, fill=blue!20] ($(R2) + (-\A/2, - 0.1)$) rectangle ($(R2) + (\A/2, 0.1)$);   
	\draw[thick, rotate=\Th, fill=yellow!20] ($(Rcm) + (0.05,-\Lone)$) rectangle  ($(Rcm) + (-0.05,\Ltwo)$);  
	\draw[thick, fill=red!40] (Rcm) circle (0.07);           
	\draw [thick, fill=blue!20] (R1) circle (0.18);                                                    
	\draw[->,thick] (R1) -- ($(R1) - (0.9,-0.1)$) node[below] {$\vec F_1$}; 
	\draw[fill] (R2) circle (0.05);               
	\end{scope}
	\begin{scope}[shift={(-0.3500,-7.5-0.3)}]
	\def\a{3}; 
	\def\lone{0.75};
	\def\ltwo{1.25};
	\def\th{0.2348*180/3.14159};
	\coordinate (rcm) at (0,0); 
	\coordinate (r1) at ($(rcm) + ({\lone*sin(\th)}, {-\lone*cos(\th)})$);
	\coordinate (r2) at ($(rcm) + ({-\ltwo*sin(\th)},{\ltwo*cos(\th)})$);
	\draw[thick, fill=blue!20] ($(r2) + (-\a/2, - 0.1)$) rectangle ($(r2) + (\a/2, 0.1)$);   
	\draw[thick, rotate=\th, fill=yellow!20] ($(rcm) + (0.05,-\lone)$) rectangle  ($(rcm) + (-0.05,\ltwo)$);  
	\draw[thick, fill=red!40] (rcm) circle (0.07);           
	\draw [thick, fill=blue!20] (r1) circle (0.18);                                                    

	\draw[fill] (r2) circle (0.05);                                                                                     
	\draw [dashed,thick] ($(rcm) + ({\lone*sin(\th)}, {-6.6})$) -- ($(rcm) + ({\lone*sin(\th)}, {2.45})$);  
	\draw[->,thick] ($({\lone*sin(\th)}, {-6.2})$)  --+ (2.6291,0) node[left, shift={(-1.5,0.05)}] {$\Delta x_1$};     
	\draw[dotted,thick] ($(r2) + (0,-5.75)$) -- ($(r2) + (0,3.1)$);    
	\draw[->,thick] ($({-\ltwo*sin(\th)},{-4.2})$)  --+ (4.5516,0) node[left, shift={(-2.7,0.05)}] {$\Delta x_2$};  
	\draw[dashdotted,thick] ($(rcm) + (0,3.3)$) -- ($(rcm) + (0,-5.6)$);   
	\draw[->,thick] (0,-5.2)  --+ (3.35,0) node[left, shift={(-1.95,0.05)}] {$\Delta x_{\rm cm}$};  
	\end{scope}
	\begin{scope}[shift={(0,-1.5-7.5-0.6)}]
	\def\A{3};
	\def\Lone{0.75};
	\def\Ltwo{1.25};
	\def\Th{3.3763*180/3.14159};
	\coordinate (Rcm) at (0,0); 
	\coordinate (R1) at ($(Rcm) + ({\Lone*sin(\Th)}, {-\Lone*cos(\Th)})$);
	\coordinate (R2) at ($(Rcm) + ({-\Ltwo*sin(\Th)},{\Ltwo*cos(\Th)})$);
	\draw[thick, fill=blue!20] ($(R2) + (-\A/2, - 0.1)$) rectangle ($(R2) + (\A/2, 0.1)$);   
	\draw[thick, rotate=\Th, fill=yellow!20] ($(Rcm) + (0.05,-\Lone)$) rectangle  ($(Rcm) + (-0.05,\Ltwo)$);  
	\draw[thick, fill=red!40] (Rcm) circle (0.07);           
	\draw [thick, fill=blue!20] (R1) circle (0.18);                                                    
	\draw[fill] (R2) circle (0.05);               
	\end{scope}
	\begin{scope}[shift={(3.0,-15-0.6)}]
	\def\a{3}; 
	\def\lone{0.75};
	\def\ltwo{1.25};
	\def\th{-0.8149*180/3.14159};
	\coordinate (rcm) at (0,0); 
	\coordinate (r1) at ($(rcm) + ({\lone*sin(\th)}, {-\lone*cos(\th)})$);
	\coordinate (r2) at ($(rcm) + ({-\ltwo*sin(\th)},{\ltwo*cos(\th)})$);
	\draw[thick, fill=blue!20] ($(r2) + (-\a/2, - 0.1)$) rectangle ($(r2) + (\a/2, 0.1)$);   
	\draw[thick, rotate=\th, fill=yellow!20] ($(rcm) + (0.05,-\lone)$) rectangle  ($(rcm) + (-0.05,\ltwo)$);  
	\draw[thick, fill=red!40] (rcm) circle (0.07);           
	\draw [thick, fill=blue!20] (r1) circle (0.18);                                                    
	\draw[->,thick] (r1) -- ($(r1) - (0.85,0.35)$) node[above] {$\vec f_1$}; 
	\draw[fill] (r2) circle (0.05);                                                                                     
	\draw [dashed,thick] ($(rcm) + ({\lone*sin(\th)}, {-0.4})$) -- ($(rcm) + ({\lone*sin(\th)}, {2.15})$);  
	\draw[dotted,thick] ($(r2) + (0,0)$) -- ($(r2) + (0,3.2)$);    
	\draw[dashdotted,thick] ($(rcm) + (0,3.2)$) -- ($(rcm) + (0,0)$);   
	\end{scope}
	\begin{scope}[shift={(0,-1.5-15-0.9)}]
	\def\A{3};
	\def\Lone{0.75};
	\def\Ltwo{1.25};
	\def\Th{2.3267*180/3.14159};
	\coordinate (Rcm) at (0,0); 
	\coordinate (R1) at ($(Rcm) + ({\Lone*sin(\Th)}, {-\Lone*cos(\Th)})$);
	\coordinate (R2) at ($(Rcm) + ({-\Ltwo*sin(\Th)},{\Ltwo*cos(\Th)})$);
	\draw[thick, fill=blue!20] ($(R2) + (-\A/2, - 0.1)$) rectangle ($(R2) + (\A/2, 0.1)$);   
	\draw[thick, rotate=\Th, fill=yellow!20] ($(Rcm) + (0.05,-\Lone)$) rectangle  ($(Rcm) + (-0.05,\Ltwo)$);  
	\draw[thick, fill=red!40] (Rcm) circle (0.07);           
	\draw [thick, fill=blue!20] (R1) circle (0.18);                                                    
	\draw[->,thick] (R1) -- ($(R1) + (0.85,0.35)$) node[below] {$\vec F_1$}; 
	\draw[fill] (R2) circle (0.05);               
	\end{scope}
\end{tikzpicture}
\caption{Forced sliding with attractive electric interaction. The asperites are shown at times $t = 15.0$ (top), $19.3$ (middle) and $26.0$ (bottom).}
\label{ForcedAttractiveFig}
\end{figure}
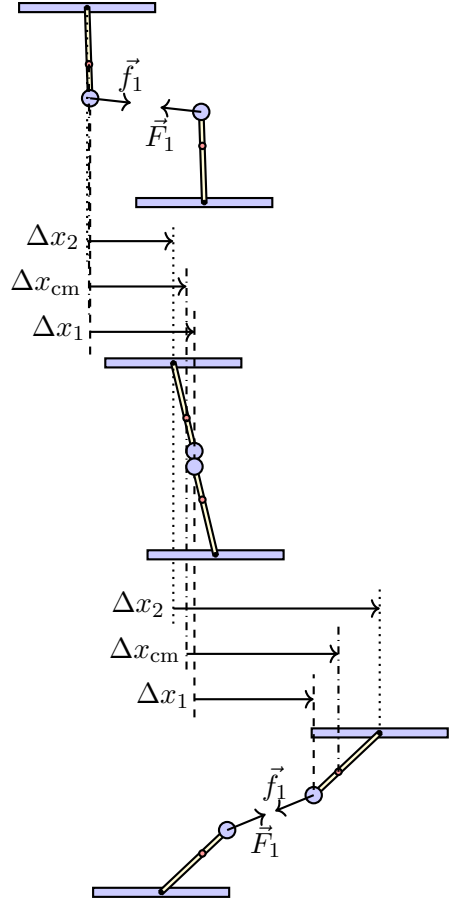
The displacements of the pendulum bob ($\Delta x_1$),  bar  ($\Delta x_2$) and center of mass ($\Delta x_{\rm cm}$) are indicated for both the approach and recede phases. 
These displacements are all positive.  

During the approach phase, $f_{1,x}$  is positive while $f_{2,x}$ is negative. Because the asperites are flexible, the attractive interaction pulls 
the pendulum bob  
ahead of the center of mass, while $\vec f_2$ pushes the bar behind.  Therefore  $\Delta x_1 > \Delta x_{\rm cm} > \Delta x_2$. The work and pseudowork 
done by friction $\vec f_1$ are positive, so that $w_1 > w_{1,{\rm ps}}$. The work and pseudowork done by the applied force $\vec f_2$ are 
negative, so that $w_2 > w_{2,{\rm ps}}$. 

During the recede phase, $f_{1,x}$  is negative while $f_{2,x}$ is positive. The attractive interaction pulls 
the pendulum bob  
behind the center of mass, while $\vec f_2$ pushes the bar forward.  Therefore  $\Delta x_2 > \Delta x_{\rm cm} > \Delta x_1$. The work and pseudowork 
done by friction $\vec f_1$ are negative, so that $w_1 > w_{1,{\rm ps}}$. The work and pseudowork done by the applied force $\vec f_2$ are 
positive, so that $w_2 > w_{2,{\rm ps}}$. 

Once again we see that the work done by each force on the flexible asperite is greater than the pseudowork. This can be seen in 
Fig.~\ref{workpowerEAB}, where the pseudowork curves lie below the corresponding work curves. Therefore $w > w_{\rm ps}$, and 
the internal energy of the asperite increases. 

Since $w_{\rm ps} =0$ for forced sliding, we can also conclude that $w>0$. 
As in the previous subsection,  the reasoning here is oversimplified because it implies that the net work always increases.  
From the graph of Fig.~\ref{workpowerEAB} we see that the net work actually decreases for a brief time period from $t=19.2$ to $19.9$. 

\subsection{Free Sliding with repulsive interaction}\label{Sec:resultsfree}
Now consider free sliding with electric repulsion, $qQ/(4\pi\epsilon_0) = 1/5$. The lower asperite is stationary. 
 The applied forces on the upper asperite are modified as described in Sec.~\ref{Sec:forces}  by setting the $x$--component of $\vec f_2$ to zero.  
The parameter values are now 
$m_1 = M_1 = 4$,  $m_2 = M_2 = 8$,  $\kappa = K =  2$, $\ell = L = 2$, $a = A = 3$, and $i_2 = I_2= 5$.  

\begin{figure}[htb]
\centering
\includegraphics[scale=1]{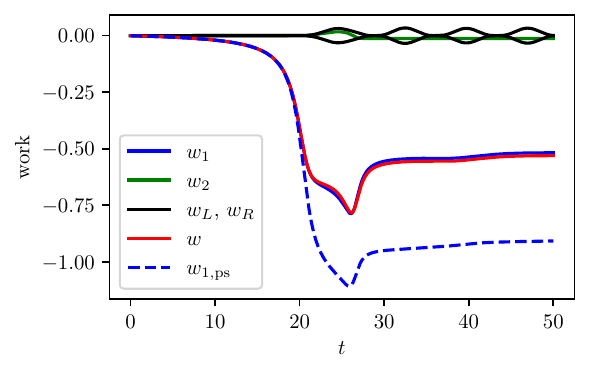}
\caption{Work done on the upper asperite for free sliding with repulsive interaction. }
\label{WorkPowerERU}
\end{figure}

The work done on the upper asperite is shown in 
Fig.~\ref{WorkPowerERU}. Note that the work done by applied forces is nearly zero. In particular, the small 
amount of work done by $\vec f_2$ is due to the small motion of the bar in the $y$--direction, and the work done by $\vec f_L$ and $\vec f_R$ 
cancel.  The net work is primarily due to friction. 

At the end of the simulation we find  $w_1 = -0.518$ and $w_2 = -0.013$. The net 
work is negative, $w = -0.531$. The translational kinetic energy decreases, $\Delta k_{\rm trans} = -0.908$, while 
the internal energy increases, $\Delta e_{\rm int} = 0.377$. The change in total energy, $\Delta k_{\rm trans} + \Delta e_{\rm int} = -0.531$,
agrees with the net work $w$, as required by the work--energy principle (\ref{wetheorem}). 

The pseudowork $w_{1,{\rm ps}}$ delivered by friction  is  
shown as a dashed curve in Fig.~\ref{WorkPowerERU}.   For the applied force $\vec f_2$, the  pseudowork vanishes ($w_{2,{\rm ps}} = 0$). 
The pseudowork at the end of the simulation, $w_{1,{\rm ps}} = -0.908$,  agrees with the change in 
translation kinetic energy as required by the pseudowork--energy principle Eq.~(\ref{psworkenergythm}). 

To understand why work is greater than 
pseudowork in this example, we first divide the 
motion into an approach phase and a recede phase. The pendulum bobs pass each other  at $t = 25.9$, when $w_1$ reaches a minimum (the 
power delivered by friction switches from negative to positive). Before 
$t \approx 13$ and after $t \approx 30$ the work done is insignificant. 
We define the approach phase  by $13.0 \le t \le 25.9$ and the recede phase by $25.9 \le t \le 30.0$. 
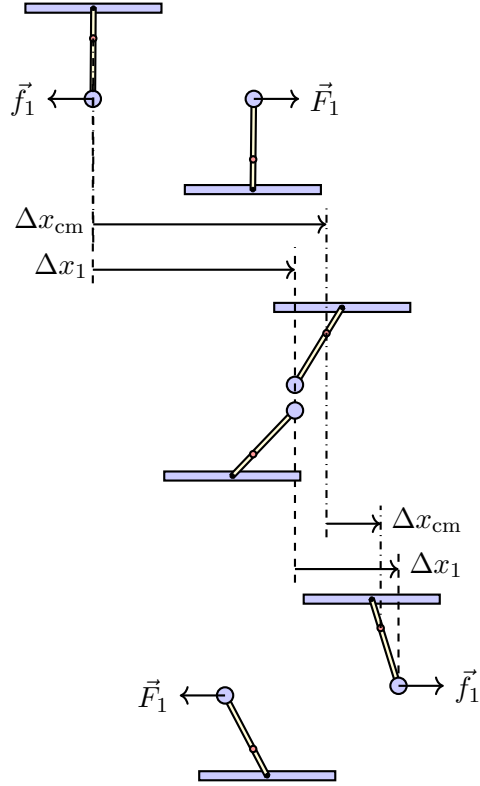
\begin{figure}[htb]
\begin{tikzpicture}[scale=0.6]
	\begin{scope}[shift={(-3.5265,0)}]
	\def\a{3}; 
	\def\lone{1.3333};
	\def\ltwo{0.6667};
	\def\th{-0.0068*180/3.14159};
	\coordinate (rcm) at (0,0); 
	\coordinate (r1) at ($(rcm) + ({\lone*sin(\th)}, {-\lone*cos(\th)})$);
	\coordinate (r2) at ($(rcm) + ({-\ltwo*sin(\th)},{\ltwo*cos(\th)})$);
	\draw[thick, fill=blue!20] ($(r2) + (-\a/2, - 0.1)$) rectangle ($(r2) + (\a/2, 0.1)$);   
	\draw[thick, rotate=\th, fill=yellow!20] ($(rcm) + (0.05,-\lone)$) rectangle  ($(rcm) + (-0.05,\ltwo)$);  
	\draw[thick, fill=red!40] (rcm) circle (0.07);           
	\draw [thick, fill=blue!20] (r1) circle (0.18);                                                    
	\draw[->,thick] (r1) -- ($(r1) - (1,0)$) node[left] {$\vec f_1$}; 
	\draw[fill] (r2) circle (0.05);                                                                                     
	\draw [dashed,thick] ($(rcm) + ({\lone*sin(\th)}, {-5.4})$) -- ($(rcm) + ({\lone*sin(\th)}, {-1.0})$);  
	\draw[->,thick] ($({\lone*sin(\th)}, {-5.1})$)  --+ (4.4561,0) node[left, shift={(-2.6,0.05)}] {$\Delta x_1$};     
	\draw[dashdotted,thick] (rcm) -- ($(rcm) + (0,-4.5)$);   
	\draw[->,thick] ($({0},{-4.1})$)  --+ (5.1416,0) node[left, shift={(-3.05,0.05)}] {$\Delta x_{\rm cm}$};  
	\end{scope}
	\begin{scope}[shift={(0,-2.6667)}]   
	\def\A{3};
	\def\Lone{1.3333};
	\def\Ltwo{0.6667};
	\def\Th{3.1313*180/3.14159};
	\coordinate (Rcm) at (0,0); 
	\coordinate (R1) at ($(Rcm) + ({\Lone*sin(\Th)}, {-\Lone*cos(\Th)})$);
	\coordinate (R2) at ($(Rcm) + ({-\Ltwo*sin(\Th)},{\Ltwo*cos(\Th)})$);
	\draw[thick, fill=blue!20] ($(R2) + (-\A/2, - 0.1)$) rectangle ($(R2) + (\A/2, 0.1)$);   
	\draw[thick, rotate=\Th, fill=yellow!20] ($(Rcm) + (0.05,-\Lone)$) rectangle  ($(Rcm) + (-0.05,\Ltwo)$);  
	\draw[thick, fill=red!40] (Rcm) circle (0.07);           
	\draw [thick, fill=blue!20] (R1) circle (0.18);                                                    
	\draw[->,thick] (R1) -- ($(R1) + (1,0)$) node[right] {$\vec F_1$};
	\draw[fill] (R2) circle (0.05);               
	\end{scope}
	\begin{scope}[shift={(1.6163,-6.5)}]
	\def\a{3}; 
	\def\lone{1.3333};
	\def\ltwo{0.6667};
	\def\th{-0.5484*180/3.14159};
	\coordinate (rcm) at (0,0); 
	\coordinate (r1) at ($(rcm) + ({\lone*sin(\th)}, {-\lone*cos(\th)})$);
	\coordinate (r2) at ($(rcm) + ({-\ltwo*sin(\th)},{\ltwo*cos(\th)})$);
	\draw[thick, fill=blue!20] ($(r2) + (-\a/2, - 0.1)$) rectangle ($(r2) + (\a/2, 0.1)$);   
	\draw[thick, rotate=\th, fill=yellow!20] ($(rcm) + (0.05,-\lone)$) rectangle  ($(rcm) + (-0.05,\ltwo)$);  
	\draw[thick, fill=red!40] (rcm) circle (0.07);           
	\draw [thick, fill=blue!20] (r1) circle (0.18);                                                    
	\draw[fill] (r2) circle (0.05);                                                                                     
	\draw [dashed,thick] ($(rcm) + ({\lone*sin(\th)}, {-5.5})$) -- ($(rcm) + ({\lone*sin(\th)}, {1.9})$);  
	\draw[->,thick] ($({\lone*sin(\th)}, {-5.2})$)  --+ (2.2845,0) node[right, shift={(0.0,0.05)}] {$\Delta x_1$};     
	\draw[dashdotted,thick] ($(rcm) + (0,-4.5)$) -- ($(rcm) + (0,2.8)$);    
	\draw[->,thick] ($({0},{-4.2})$)  --+ (1.1946,0) node[right, shift={(0.0,0.05)}] {$\Delta x_{\rm cm}$};  
	\end{scope}
	\begin{scope}[shift={(0,-2.6667-6.5)}]
	\def\A{3};
	\def\Lone{1.3333};
	\def\Ltwo{0.6667};
	\def\Th{2.3781*180/3.14159};
	\coordinate (Rcm) at (0,0); 
	\coordinate (R1) at ($(Rcm) + ({\Lone*sin(\Th)}, {-\Lone*cos(\Th)})$);
	\coordinate (R2) at ($(Rcm) + ({-\Ltwo*sin(\Th)},{\Ltwo*cos(\Th)})$);
	\draw[thick, fill=blue!20] ($(R2) + (-\A/2, - 0.1)$) rectangle ($(R2) + (\A/2, 0.1)$);   
	\draw[thick, rotate=\Th, fill=yellow!20] ($(Rcm) + (0.05,-\Lone)$) rectangle  ($(Rcm) + (-0.05,\Ltwo)$);  
	\draw[thick, fill=red!40] (Rcm) circle (0.07);           
	\draw [thick, fill=blue!20] (R1) circle (0.18);                                                    
	\draw[fill] (R2) circle (0.05);               
	\end{scope}
	\begin{scope}[shift={(2.8106,-13)}]
	\def\a{3}; 
	\def\lone{1.3333};
	\def\ltwo{0.6667};
	\def\th{0.2997*180/3.14159};
	\coordinate (rcm) at (0,0); 
	\coordinate (r1) at ($(rcm) + ({\lone*sin(\th)}, {-\lone*cos(\th)})$);
	\coordinate (r2) at ($(rcm) + ({-\ltwo*sin(\th)},{\ltwo*cos(\th)})$);
	\draw[thick, fill=blue!20] ($(r2) + (-\a/2, - 0.1)$) rectangle ($(r2) + (\a/2, 0.1)$);   
	\draw[thick, rotate=\th, fill=yellow!20] ($(rcm) + (0.05,-\lone)$) rectangle  ($(rcm) + (-0.05,\ltwo)$);  
	\draw[thick, fill=red!40] (rcm) circle (0.07);           
	\draw [thick, fill=blue!20] (r1) circle (0.18);                                                    
	\draw[->,thick] (r1) -- ($(r1) + (1,0)$) node[right] {$\vec f_1$}; 
	\draw[fill] (r2) circle (0.05);                                                                                     
	\draw [dashed,thick] ($(rcm) + ({\lone*sin(\th)}, {-1.0})$) -- ($(rcm) + ({\lone*sin(\th)}, {1.8})$);  
	\draw[dashdotted,thick] ($(rcm) + (0,0)$) -- ($(rcm) + (0,2.7)$);    
	\end{scope}
	\begin{scope}[shift={(0,-2.6667-13)}]
	\def\A{3};
	\def\Lone{1.3333};
	\def\Ltwo{0.6667};
	\def\Th{3.6251*180/3.14159};
	\coordinate (Rcm) at (0,0); 
	\coordinate (R1) at ($(Rcm) + ({\Lone*sin(\Th)}, {-\Lone*cos(\Th)})$);
	\coordinate (R2) at ($(Rcm) + ({-\Ltwo*sin(\Th)},{\Ltwo*cos(\Th)})$);
	\draw[thick, fill=blue!20] ($(R2) + (-\A/2, - 0.1)$) rectangle ($(R2) + (\A/2, 0.1)$);   
	\draw[thick, rotate=\Th, fill=yellow!20] ($(Rcm) + (0.05,-\Lone)$) rectangle  ($(Rcm) + (-0.05,\Ltwo)$);  
	\draw[thick, fill=red!40] (Rcm) circle (0.07);           
	\draw [thick, fill=blue!20] (R1) circle (0.18);                                                    
	\draw[->,thick] (R1) -- ($(R1) - (1,0)$) node[left] {$\vec F_1$};
	\draw[fill] (R2) circle (0.05);               
	\end{scope}
\end{tikzpicture}
\caption{Free sliding with repulsive interaction. Asperites are shown at times $t = 13.0$ (top), $25.9$ (middle) and $30.0$ (bottom). }
\label{FreeAttractiveFig}
\end{figure}

The asperites are shown in Fig.~\ref{FreeAttractiveFig} at times $t=13.0$, $25.9$ and $30.0$. During the approach phase,  
the upper asperite's center of mass 
displacement is $\Delta x_{\rm cm} \approx 5.14$. The pendulum bob lags behind due to the repulsive 
electric force, so it moves a  shorter distance, $\Delta x_1 \approx 4.46$. 
Since the force $\vec f_1$ is primarily in the $-x$ direction during the approach, 
the power $w_1 \approx f_{1,x} \Delta x_1$ and pseudopower $w_{1,{\rm ps}} \approx f_{1,x} \Delta x_{\rm cm}$ are  negative. 
Therefore $w_1 > w_{1,{\rm ps}}$ during the approach phase. 

For the recede phase, the asperite's center of mass displacement is $\Delta x_{\rm cm} = 1.19$. The repulsive 
force pushes the pendulum bob ahead, giving a  relatively large displacement $\Delta x_1 = 2.28$. The power and pseudopower are both 
positive in this phase, so that  $w_1 > w_{1,{\rm ps}}$. 

In both the approach and recede phases, the work done by friction is greater than the pseudowork. 
 The difference between the net work and net pseudowork is $w_1 + w_2  - w_{1,{\rm ps}} - w_{2,{\rm ps}} = 0.377$. The agrees with  
 the change in internal energy, 
 $\Delta e_{\rm int} = 0.377$, as required by the IFLT.
 
 The pseudowork--energy principle (\ref{psworkenergythm}) tells us that 
 for free sliding on a stationary surface, the pseudowork should be negative. This is why the asperite loses translational kinetic energy and slows down. Observe that the 
 pseudopower delivered by friction is negative in the approach phase, and positive in the recede phase. That is, from Fig.~\ref{WorkPowerERU} we see that 
 the pseudowork $w_{1,{\rm ps}}$ has negative slope in the approach phase and positive slope in the recede phase. 
However, the magnitude of the pseudowork done during the approach phase is greater than the 
magnitude of the pseudowork done during the recede phase, so by the end of the interaction the pseudowork is negative. 

Figure \ref{FreeAttractiveFig} shows why the accumulated pseudowork is negative.  During the approach phase, the 
pendulums shift away from one another as they approach. This phase lasts a long time, which allows the upper asperite's center of mass to move a  large 
distance ($\Delta x_{\rm cm} \approx 5.14$) while the bobs are near each other and interacting strongly. 
The magnitude of  the pseudowork done during this phase is relatively large. During the recede phase, 
the pendulums shift away from one another.  This phase is brief. The  upper asperite's center of mass moves a relatively short distance 
($\Delta x_{\rm cm} \approx 1.19$) 
while the bobs are near each other. Therefore, during this phase, the magnitude of the pseudowork is relatively small. 
Overall, the pseudowork is negative because the approach--phase pseudowork (which is negative) dominates over the recede--phase pseudowork (which is positive). 

\subsection{Free Sliding with attractive interaction}\label{Sec:resultsfreeattract}
We now reverse the sign of the electric force by setting $qQ = -1/5$, and add a gap $s = 0.3$ between the 
asperites. Otherwise, the parameters and initial conditions are the same as in the previous subsection. 

\begin{figure}[htb]
\centering
\includegraphics[scale=1]{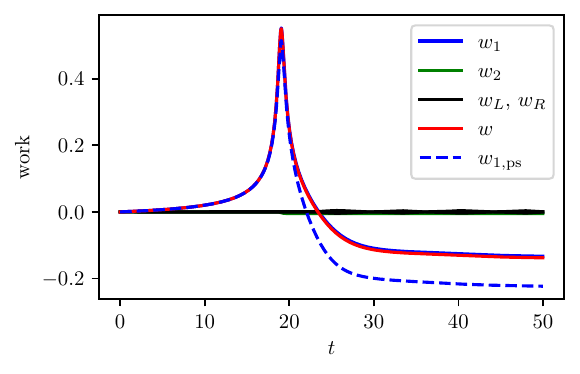}
\caption{Work done on the upper asperite for free sliding with attractive interaction. }
\label{WorkPowerEAU}
\end{figure}

Figure \ref{WorkPowerEAU} shows the work delivered to the upper asperite. By the end of the simulation, the work done by applied 
forces is very small ($w_2 = -0.00486$, $w_L + w_R = 0$).  The net work is almost entirely due to friction, $w_1 = -0.133$. 

The dashed curve in Fig.~\ref{WorkPowerEAU} is the  pseudowork done by friction. 
The  pseudowork vanishes for the applied forces; in particular, $w_{2,{\rm ps}} = 0$. By the end of the simulation the pseudowork supplied  by 
friction is $w_{1,{\rm ps}} = -0.223$. 

Overall, the real work ($w_1 + w_2 = -0.138$) is greater than the pseudowork ($w_{1,{\rm ps}} + w_{2,{\rm ps}} = -0.223$). 
The difference  equals the change in internal energy, $\Delta e_{\rm int} = 0.085$, as required by the IFLT. 
The pseudowork itself  is equal to the change in translational kinetic energy of the asperite, $\Delta k_{\rm trans} = -0.223$, 
as required by the pseudowork--energy principle. 

Why is work greater than pseudowork? Since the force is attractive the power and pseudopower are  positive as the 
pendulum bobs approach and negative as they recede. Due to the flexibility of the asperites, the displacement of the pendulum bob 
during the approach phase 
is greater than the displacement of the center of mass. Therefore, during the approach, work is greater than pseudowork. As the 
pendulum bobs recede, the power and pseudopower are  negative. The upper pendulum shifts back  so that the 
center of mass displacement is greater than the bob displacement. Once again, work is greater than pseudowork. 

Why is  pseudowork negative?
As the pendulum bobs approach,  the attractive force 
pulls the lower bob to the left and the upper bob to the right. The center of mass of the upper asperite moves a relatively small 
distance during the brief time in which the bobs are in close proximity to one another. Therefore, as the bobs approach, the pseudowork done by friction 
is positive but relatively small. 
As the pendulum bobs recede, the attractive force pulls the lower bob to the right and the upper bob to the left. The center of mass of the 
upper asperite moves a relatively large distance while the bobs are in close proximity. Thus, in this phase, the pseudowork 
done by friction is negative and has a relatively large magnitude. 
Overall, the pseudowork done by friction is negative. By the pseudowork--energy principle (\ref{psworkenergythm}) the translational kinetic 
energy of the upper asperite decreases. 

\subsection{Free sliding from rest on a moving table}\label{Sec:referenceframe}
Consider a block, initially at rest, dropped onto a moving table. The block slides freely thereafter.  This situation is physically equivalent  to the 
cases considered in subsecs.~\ref{Sec:resultsfree} and \ref{Sec:resultsfreeattract}, but viewed in a reference frame in which the block (upper asperite) is initially at rest. In the 
model code, we simply change the initial conditions from $\vec v_{\rm cm} = \langle 1/2,0,0 \rangle$, $\vec V_{\rm cm} = \langle 0,0,0 \rangle$ 
to $\vec v_{\rm cm} = \langle 0,0,0 \rangle$, $\vec V_{\rm cm} = \langle -1/2,0,0 \rangle$. 

Work and pseudowork depend on the frame of reference, but the difference between work and pseudowork is frame independent. By the IFLT,
this difference equals the (frame independent) change in internal energy. Whether we view the block--table interaction in the ``old" 
reference frame, with the table at rest, or in the ``new" reference frame, with the block initially at rest, the change in internal energy is the same. 

On the other hand,  the pseudowork done on a system and the translational kinetic energy of the system do 
depend on the frame of reference. 
We can modify the results of subsecs.~\ref{Sec:resultsfree} and \ref{Sec:resultsfreeattract} to reflect a change  from the ``old"  frame 
to the ``new" frame  by  simply applying the coordinate transformation
$x_{\rm new} = x_{\rm old} - v_it$. Here,  $v_i = 1/2$ is the initial velocity of the upper asperite (in the old frame). 
With this change, the displacements for the upper 
asperite become $\Delta x_{\rm new} = \Delta x_{\rm old} - v_i\Delta t$. 

\begin{figure}[htb]
\begin{tikzpicture}[scale=0.6]
	\begin{scope}[shift={(-3.5265,0)}]
	\def\a{3}; 
	\def\lone{1.3333};
	\def\ltwo{0.6667};
	\def\th{-0.0068*180/3.14159};
	\coordinate (rcm) at (0,0); 
	\coordinate (r1) at ($(rcm) + ({\lone*sin(\th)}, {-\lone*cos(\th)})$);
	\coordinate (r2) at ($(rcm) + ({-\ltwo*sin(\th)},{\ltwo*cos(\th)})$);
	\draw[thick, fill=blue!20] ($(r2) + (-\a/2, - 0.1)$) rectangle ($(r2) + (\a/2, 0.1)$);   
	\draw[thick, rotate=\th, fill=yellow!20] ($(rcm) + (0.05,-\lone)$) rectangle  ($(rcm) + (-0.05,\ltwo)$);  
	\draw[thick, fill=red!40] (rcm) circle (0.07);           
	\draw [thick, fill=blue!20] (r1) circle (0.18);                                                    
	\draw[->,thick] (r1) -- ($(r1) - (1,0)$) node[left] {$\vec f_1$}; 
	\draw[fill] (r2) circle (0.05);                                                                                     
	\draw [dashed,thick] ($(rcm) + ({\lone*sin(\th)}, {-5.4})$) -- ($(rcm) + ({\lone*sin(\th)}, {-1.0})$);  
	\draw[->,thick] ($({\lone*sin(\th)}, {-5.1})$)  --+ (4.4561 - 6.435,0) node[left, shift={(0,0.05)}] {$\Delta x_1$};     
	\draw[dashdotted,thick] (rcm) -- ($(rcm) + (0,-4.5)$);   
	\draw[->,thick] ($({0},{-4.1})$)  --+ (5.1416 - 6.435,0) node[left, shift={(0,0.05)}] {$\Delta x_{\rm cm}$};  
	\end{scope}
	\begin{scope}[shift={(0,-2.6667)}]   
	\def\A{3};
	\def\Lone{1.3333};
	\def\Ltwo{0.6667};
	\def\Th{3.1313*180/3.14159};
	\coordinate (Rcm) at (0,0); 
	\coordinate (R1) at ($(Rcm) + ({\Lone*sin(\Th)}, {-\Lone*cos(\Th)})$);
	\coordinate (R2) at ($(Rcm) + ({-\Ltwo*sin(\Th)},{\Ltwo*cos(\Th)})$);
	\draw[thick, fill=blue!20] ($(R2) + (-\A/2, - 0.1)$) rectangle ($(R2) + (\A/2, 0.1)$);   
	\draw[thick, rotate=\Th, fill=yellow!20] ($(Rcm) + (0.05,-\Lone)$) rectangle  ($(Rcm) + (-0.05,\Ltwo)$);  
	\draw[thick, fill=red!40] (Rcm) circle (0.07);           
	\draw [thick, fill=blue!20] (R1) circle (0.18);                                                    
	\draw[->,thick] (R1) -- ($(R1) + (1,0)$) node[right] {$\vec F_1$}; 
	\draw[fill] (R2) circle (0.05);               
	\end{scope}
	\begin{scope}[shift={(1.6163 - 6.435,-6.5)}]
	\def\a{3}; 
	\def\lone{1.3333};
	\def\ltwo{0.6667};
	\def\th{-0.5484*180/3.14159};
	\coordinate (rcm) at (0,0); 
	\coordinate (r1) at ($(rcm) + ({\lone*sin(\th)}, {-\lone*cos(\th)})$);
	\coordinate (r2) at ($(rcm) + ({-\ltwo*sin(\th)},{\ltwo*cos(\th)})$);
	\draw[thick, fill=blue!20] ($(r2) + (-\a/2, - 0.1)$) rectangle ($(r2) + (\a/2, 0.1)$);   
	\draw[thick, rotate=\th, fill=yellow!20] ($(rcm) + (0.05,-\lone)$) rectangle  ($(rcm) + (-0.05,\ltwo)$);  
	\draw[thick, fill=red!40] (rcm) circle (0.07);           
	\draw [thick, fill=blue!20] (r1) circle (0.18);                                                    
	\draw[fill] (r2) circle (0.05);                                                                                     
	\draw [dashed,thick] ($(rcm) + ({\lone*sin(\th)}, {-5.5})$) -- ($(rcm) + ({\lone*sin(\th)}, {1.9})$);  
	\draw[->,thick] ($({\lone*sin(\th)}, {-5.2})$)  --+ (2.2845 - 2.065,0) node[right, shift={(0.0,0.05)}] {$\Delta x_1$};     
	\draw[dashdotted,thick] ($(rcm) + (0,-4.5)$) -- ($(rcm) + (0,2.8)$);    
	\draw[->,thick] ($({0},{-4.2})$)  --+ (1.1946 - 2.065,0) node[left, shift={(0.0,0.05)}] {$\Delta x_{\rm cm}$};  
	\end{scope}
	\begin{scope}[shift={(0- 6.435,-2.6667-6.5)}]
	\def\A{3};
	\def\Lone{1.3333};
	\def\Ltwo{0.6667};
	\def\Th{2.3781*180/3.14159};
	\coordinate (Rcm) at (0,0); 
	\coordinate (R1) at ($(Rcm) + ({\Lone*sin(\Th)}, {-\Lone*cos(\Th)})$);
	\coordinate (R2) at ($(Rcm) + ({-\Ltwo*sin(\Th)},{\Ltwo*cos(\Th)})$);
	\draw[thick, fill=blue!20] ($(R2) + (-\A/2, - 0.1)$) rectangle ($(R2) + (\A/2, 0.1)$);   
	\draw[thick, rotate=\Th, fill=yellow!20] ($(Rcm) + (0.05,-\Lone)$) rectangle  ($(Rcm) + (-0.05,\Ltwo)$);  
	\draw[thick, fill=red!40] (Rcm) circle (0.07);           
	\draw [thick, fill=blue!20] (R1) circle (0.18);                                                    
	\draw[fill] (R2) circle (0.05);               
	\end{scope}
	\begin{scope}[shift={(2.8106 - 8.5,-13)}]
	\def\a{3}; 
	\def\lone{1.3333};
	\def\ltwo{0.6667};
	\def\th{0.2997*180/3.14159};
	\coordinate (rcm) at (0,0); 
	\coordinate (r1) at ($(rcm) + ({\lone*sin(\th)}, {-\lone*cos(\th)})$);
	\coordinate (r2) at ($(rcm) + ({-\ltwo*sin(\th)},{\ltwo*cos(\th)})$);
	\draw[thick, fill=blue!20] ($(r2) + (-\a/2, - 0.1)$) rectangle ($(r2) + (\a/2, 0.1)$);   
	\draw[thick, rotate=\th, fill=yellow!20] ($(rcm) + (0.05,-\lone)$) rectangle  ($(rcm) + (-0.05,\ltwo)$);  
	\draw[thick, fill=red!40] (rcm) circle (0.07);           
	\draw [thick, fill=blue!20] (r1) circle (0.18);                                                    
	\draw[->,thick] (r1) -- ($(r1) + (1,0)$) node[right] {$\vec f_1$}; 
	\draw[fill] (r2) circle (0.05);                                                                                     
	\draw [dashed,thick] ($(rcm) + ({\lone*sin(\th)}, {-1.0})$) -- ($(rcm) + ({\lone*sin(\th)}, {1.8})$);  
	\draw[dashdotted,thick] ($(rcm) + (0,0)$) -- ($(rcm) + (0,2.7)$);    
	\end{scope}
	\begin{scope}[shift={(0 - 8.5,-2.6667-13)}]
	\def\A{3};
	\def\Lone{1.3333};
	\def\Ltwo{0.6667};
	\def\Th{3.6251*180/3.14159};
	\coordinate (Rcm) at (0,0); 
	\coordinate (R1) at ($(Rcm) + ({\Lone*sin(\Th)}, {-\Lone*cos(\Th)})$);
	\coordinate (R2) at ($(Rcm) + ({-\Ltwo*sin(\Th)},{\Ltwo*cos(\Th)})$);
	\draw[thick, fill=blue!20] ($(R2) + (-\A/2, - 0.1)$) rectangle ($(R2) + (\A/2, 0.1)$);   
	\draw[thick, rotate=\Th, fill=yellow!20] ($(Rcm) + (0.05,-\Lone)$) rectangle  ($(Rcm) + (-0.05,\Ltwo)$);  
	\draw[thick, fill=red!40] (Rcm) circle (0.07);           
	\draw [thick, fill=blue!20] (R1) circle (0.18);                                                    
	\draw[->,thick] (R1) -- ($(R1) - (1,0)$) node[left] {$\vec F_1$}; 
	\draw[fill] (R2) circle (0.05);               
	\end{scope}
\end{tikzpicture}
\caption{Free sliding with repulsive interaction, viewed from the inertial frame in which the upper asperite is initially at rest and the 
lower asperite is moving from right to left.}
\label{FreeAttractiveFigNewFrame}
\end{figure}
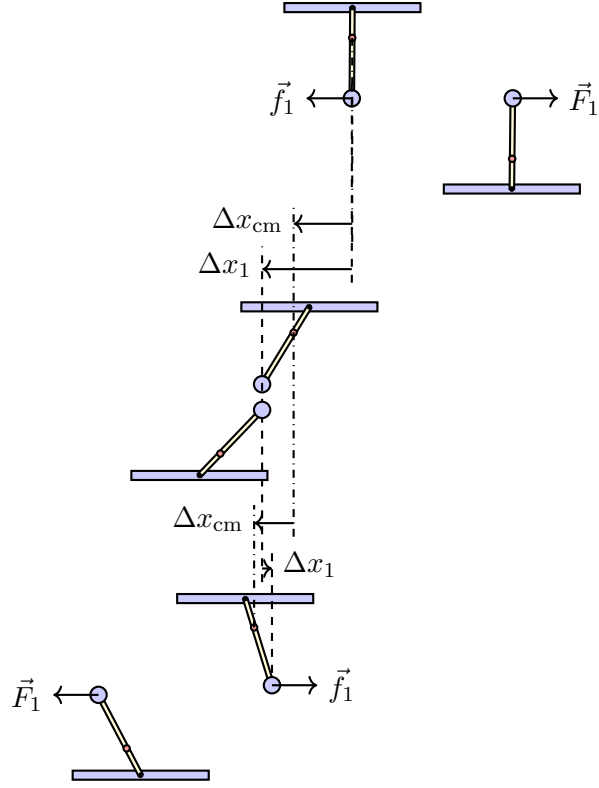

Consider the case of free sliding with a repulsive interaction, as in subsec.~\ref{Sec:resultsfree}. 
In the ``old" reference frame, the center of mass displacements  are $\Delta x_{\rm cm} = 5.14$ in the approach phase 
and $\Delta x_{\rm cm} = 1.19$ in the recede phase. The duration of the approach phase ($13.0 \le t \le 25.9$) is $\Delta t = 12.9$, 
and the duration of the recede phase ($25.9 \le t \le 30.0$) is $\Delta t = 4.1$. Therefore, in the new frame, 
the center of mass displacements are $\Delta x_{\rm cm} = 5.14 - 12.9/2 = -1.29$ and 
$\Delta x_{\rm cm} = 1.19 - 4.1/2 = -0.87$ for the approach and recede phases, respectively. 
These displacements are depicted in Fig.~\ref{FreeAttractiveFigNewFrame}. 

In the new reference frame, the approach--phase work done on the upper asperite  is positive. The repulsive force pushes the pendulum 
bobs apart, allowing this phase to last a long time ($\Delta t = 12.9$).  
The upper asperite's center of mass moves a relatively large distance during this phase, in spite of the fact that it starts from rest. During this phase the pseudowork is positive and large. 
The recede phase, on the other hand, is relatively brief ($\Delta t = 4.1$) because the pendulum bobs quickly swing away from one another. 
The center of mass moves a relatively small distance, so the pseudowork is negative and small in magnitude. 

Since the magnitude of the approach--phase pseudowork is greater than the magnitude of the recede--phase pseudowork, the net 
pseudowork is positive $w_{\rm ps} > 0$. 
By the pseudowork--energy principle Eq.~(\ref{psworkenergythm}),  the change in translational kinetic energy is positive. Viewed in the 
new reference frame, friction causes the upper asperite to speed up.  

What happens if the interaction is attractive? That is, we have free sliding with an attractive force, as in subsec.~\ref{Sec:resultsfreeattract}, 
but now viewed from a frame in which 
 the upper asperite is initially at rest and the lower asperite approaches from the right.  During the approach,  the pendulum bobs swing toward one 
 another. The center of mass of the upper asperite 
shifts to the right, so the pseudowork is positive. For a brief time after the bobs pass, the center of mass continues to move to the right. 
The friction force is opposite to the center of mass velocity, so the 
pseudowork is negative. This negative pseudowork largely cancels the positive pseudowork from  the approach phase, so up to this point the net pseudowork done is approximately zero. Subsequently, the upper asperite comes to a stop and is then  pulled to the left 
as the lower asperite recedes. During this final part of the motion, the force of friction does positive  pseudowork on the upper asperite. Overall the pseudowork 
is positive. By the pseudowork--energy principle, the translational kinetic energy  increases and the upper asperite speeds up.

\section{Discussion}\label{Sec:discuss}
In the presence of sliding friction,  a proper treatment of work and energy must take into account the flexibility of  the interacting objects. 
The mechanism  responsible for the increase in internal energy can be described in general terms as follows. When a force acting on a flexible object is in the same direction 
as the motion of the object, the force generally does positive work.  
The point of contact of the force is pushed ``forward," 
ahead of the object's center of mass, so the force does more work than pseudowork. 
When the force opposes the motion of the object, it generally does negative work. The point of contact is pushed 
``backwards" and lags behind the center of mass. 
The magnitude of the work is less than the magnitude of the pseudowork, so again,  work is greater than  pseudowork.  

The conclusion is that for each force acting on a flexible object, whether the force is attractive or repulsive or a complicated combination of both,  
work tends to be greater than pseudowork.  Therefore the net work is generally greater than the net 
pseudowork. By the invariant first law of thermodynamics, Eq.~(\ref{Dewminuswps}), the difference between  work and pseudowork 
equals the change in internal energy. In a physical object, the internal energy dissipates into thermal energy. 

It is important to recognize that dynamical systems, such as the ones explored in this paper, are time reversible. 
In other words, we could run each of the simulations backwards in time. This is equivalent to relabeling each of the graphs with time increasing 
from right to left, and reversing
the displacements  in the time sequence figures  such as Fig.~\ref{ForcedRepulsiveFig}.  The initial state would consist 
of  upper and lower asperites oscillating ``in sync." 
In the final state, the asperite pendulums would be unexcited, at their equilibrium positions with zero internal energy.
Such a dynamical process is physically  possible, but is it likely to occur? One might guess that it is 
unlikely because the initial conditions would need to be 
fine--tuned for the asperites to loose their internal energy when they interact. (And in the case of free sliding on a 
stationary surface, for the upper asperite to speed up.)
The likelihood of such a process is a statistical question that requires further study.  

On the other hand, we might simply assume that  prior to a typical encounter, the asperites have little or no internal energy. 
This is probably a good assumption 
if the time scale for the asperites to dissipate their internal energy is short compared to the timescale between encounters. In this case 
there is really only one direction that the flow of energy can go---because the internal energy is initially at (or near) a minimum, 
it can only increase or remain unchanged by an interaction.

\begin{acknowledgments}
Thanks to Ruth Chabay, Bruce Sherwood, Aaron Titus and Stephen Spicklemire for helpful comments and encouragement. 
\end{acknowledgments}


\end{document}